\documentstyle[psfig,draft]{mn}

\def\gs{\mathrel{\raise0.35ex\hbox{$\scriptstyle >$}\kern-0.6em
\lower0.40ex\hbox{{$\scriptstyle \sim$}}}}
\def\ls{\mathrel{\raise0.35ex\hbox{$\scriptstyle <$}\kern-0.6em
\lower0.40ex\hbox{{$\scriptstyle \sim$}}}}
\def\ergs {{\rm erg} \, {\rm s}^{-1}}
\def\deg {{^{\circ}}}
\def\mnras {{MNRAS}}
\def\apj {ApJ}
\def\apjs {ApJS}
\def\apjl {ApJL}
\def\aj {AJ}
\def\aaps {A\&AS}
\def\araa {ARA\&A}
\def\pasj {PASJ}

\title[The Environmental Dependance of Galaxy Colours]
      {The Las Campanas/AAT Rich Cluster Survey: II. \\
The Environmental Dependance of Galaxy Colours in Clusters at z$\sim$0.1.}

\author[Pimbblet et al.]
       {Kevin A.\ Pimbblet,$^{\! 1}$ Ian Smail,$^{\! 1}$ 
        Tadayuki Kodama,$^{\! 2,1}$ Warrick J.\ Couch,$^{\! 3}$ \and
	Alastair C.\ Edge,$^{\! 1}$ 
        Ann I.\ Zabludoff$^{4}$ \& Eileen O'Hely$^{3}$
        \vspace*{1mm}\\
        $^1$ Department of Physics, University of Durham, South Road,
        Durham, DH1 3LE, UK\\
        $^2$ Department of Astronomy, University of Tokyo, Hongo, 
        Bunkyo-ku, Tokyo, 113-0033, Japan\\
	$^3$ School of Physics, University of New South Wales, Sydney
	NSW 2052, Australia\\
	$^4$ Steward Observatory, University of Arizona, Tucson
	AZ 85721 USA}

\date{Accepted ... ; Received ... ; in original ...}

\pagerange{000--000}

\begin{document}

\maketitle

\begin{abstract}
We present a photometric investigation
of the variation in galaxy colour with environment in 11
X-ray luminous clusters at $0.07\leq z \leq 0.16$ taken from the Las
Campanas/AAT Rich Cluster Survey.  We study the properties of the
galaxy populations in individual clusters and take advantage of
the homogeneity of the sample to combine the clusters together to
investigate weaker trends in the composite sample.  We find that modal
colours of galaxies lying on the colour-magnitude relation in the
clusters become bluer by $d(B-R) / d r_p = -0.022 \pm 0.004$ from 
the cluster core out to a projected radius of $r_p = 6$\,Mpc;
further out in radius than any previous study.  
We also examine the variation in modal galaxy colour with local 
galaxy density, $\Sigma$,  
for galaxies lying close to the colour-magnitude relation and find 
that the median colour shifts bluewards by $d(B-R) / d {\rm log}_{10}(\Sigma) 
= -0.076 \pm 0.009$ with decreasing local
density across three orders of magnitude.  We show that the position of
the red envelope of galaxies in the colour-magnitude relation does not
vary as a function of projected radius or density within the clusters, suggesting
that the change in the modal colour results from an increasing fraction
of bluer galaxies within the colour-magnitude relation, rather than a
change in the colours of the {\it whole} population.  We show that this
shift in the colour-magnitude relations with projected radius
and local-density is greater than expected from the changing
morphological mix based on the local morphology--density relation.  We
therefore conclude that we are seeing a real change in 
the properties of galaxies on the colour-magnitude relation in 
the outskirts of clusters.
The simplest interpretation of this result (and similar constraints in
local clusters) is that an increasing fraction of galaxies in the lower
density regions at large radii within clusters exhibit signatures of
star formation in the recent past, signatures which
are not seen in the evolved galaxies in the highest density regions.
\end{abstract}


%
%
\section{Introduction}

Visvanathan and Sandage (1977) first noted that the more luminous
early-type galaxies (ellipticals and S0s) within Coma and eight other
local clusters exhibit systematically redder integrated colours than
the late-types.  These galaxies also demonstrate a tight correlation
between their colours and magnitudes: the colour-magnitude relationship
(CMR; e.g.\ Bower, Lucey and Ellis 1992).  Kodama \& Arimoto (1997)
show that the slope of the 
colour-magnitude relation is primarily due to the mean
stellar metallicity (as suggested by Dressler 1984), while the scatter
around the relation may be due to age effects (Kodama et al.\ 1999).

As one moves from the cores of 
clusters into the surrounding field the morphological
mix in the galaxy population shifts from almost completely passive,
early-types to one dominated by star-forming spiral galaxies. This is
the origin of the Morphology-Density relation (T--$\Sigma$; 
Dressler 1980).  The
environmental ``transition'' region between clusters and the field is
an important region to study if we are to learn what impact environment
has on galaxy properties such as star formation rate, luminosity and
morphology.  Recent numerical simulations of clusters of galaxies show
that the history of galaxy accretion is retained in the cluster's
radial profile.  The first galaxies to be accreted still reside
nearer the centre while more recent additions are located
preferentially on the outskirts of clusters (Diaferio et al.\ 2001).
This means that  
if the star formation in galaxies is suppressed when they enter the
cluster environment (Balogh, Navarro \& Morris 2000), then the
variation in the typical accretion age of cluster galaxies with radius
may be visible in the ages of the stellar populations of the passive
cluster members which inhabit the colour-magnitude relation.  Studies
of galaxies in the outskirts of clusters and the variation in their
properties with redshift can therefore inform us about whether environment
has altered the characteristics of cluster galaxies as the clusters
have been assembled over the Hubble time.

There are two main observational techniques to trace the variation in
the stellar populations of galaxies between the field and cluster and
its evolution with redshift: photometry and spectroscopy.

Wide-field photometric studies of clusters have uncovered evidence of
changes in the colours of the galaxy populations with radius: Butcher
\& Oemler (1978, 1984) first identified a photometric radial gradient;
the fraction of blue galaxies increases with projected distance, out
to $\sim 1.5$\,Mpc\footnote{Throughout this work values of $H_0 = 50
$ km s$^{-1}$\,Mpc$^{-1}$ and $q_o=0.5$ have been adopted.} (as well
as increasing dramatically with redshift).  Kodama \& Bower (2000)
confirm this strong trend in their recent photometric study of the
intermediate redshift clusters from the CNOC survey (Yee et al.\ 1996).
Such trends of average colour with environment (i.e.\ radius or 
local density) could be due to (i) a systematic, intrinsic 
cluster colour gradient in one or both of the dominant
elliptical and S0 populations of the CMR with environment; or (ii)
be primarily driven by the morphology-density relation
(Dressler 1980), where the late-type to early-type galaxies ratio is a
strongly decreasing function of local galaxy density, or equivalently
an increasing function of clustocentric distance (Whitmore \& Gilmore
1991).  Thus the observed trend may simply reflect evolution
in the morphological mix, rather than representing changes
in the colours of the galaxies lying on the CMR with environment
(as we discuss below, variations in the morphological mix
{\it within} the CMR may produce more subtle changes in colour 
with environment).

To reliably differentiate between these options 
we need to compare the properties of
morphologically-selected (or, failing, that colour-selected) early-type
galaxies in the outskirts of clusters with similar galaxies in the high
density cores.  
In this way we can determine if the blueing trend occurs in the
passive cluster population or simply reflects an increasing 
proportion of late-type galaxies.

Unfortunately, very few of the existing studies of clusters have
the field of view, photometric precision (the expected colour shifts
are $\ls 0.10$ mags) or colour information necessary to
investigate the possible variation with environment of the colour of
the galaxies on the CMR.  One of the few relevant studies is the recent
investigation by Terlevich et al.\ (2001) of the $(U-V)$ colours of
$\sim 100$ morphologically-classified early-type galaxies across a
1.5$\deg$ field within the Coma cluster ($z=0.023$).  They identify a
trend for the colours of early-type galaxies to be
systematically bluer at a fixed luminosity outside the core of
the cluster. They interpret this as a result of age differences and
suggest that there is a 2\,Gyr difference between the mean
luminosity-weighted ages of the stellar populations of galaxies in the
core and those outside 0.5\,Mpc.

%
%
\begin{table*}
\begin{center}
\medskip
\caption{\small{Details of the clusters from the LARCS sample used in
this work.  For each cluster we give the coordinates of the cluster
centre, the redshift ($z$) and its X-ray luminosity, $L_X$, in the
0.1--2.4\,keV passband (Ebeling et al.\ 1996).  
We also list the parameters of the biweight
fit to the CMR within 2\,Mpc of the centre of each of the clusters as
plotted in Figure~\ref{fig:inners}.  The slope, colour at the fiducial
magnitude and associated errors are taken from the median of 100
realizations of the background subtraction.  
These errors may be underestimated due to the way in which
the CMR is fitted (see text for details).
The apparent $R$-band
magnitude corresponding to our fiducial absolute magnitude  of
$M_V=-21.8$ is also given for each cluster.  The colour of each CMR at
a fiducial magnitude equivalent to $M_V=-21.8$ and the slope of the
relationship are examined as a function of redshift in Figure~\ref{fig:ZvsC}.
}}
\begin{tabular}{lcccccccc}
\hline
Cluster & R.A.\ & Dec.\ & $z$ & $L_X$ & Slope & $(B-R)_{M_V=-21.8}$ & $R_{M_V=-21.8}$	 \\
 & \multispan2{\hfil (J2000) \hfil } &  & ($10^{44}\ergs$) & & & \\
\hline\hline
A\,22    & 00 20 38.64 &  $-$25 43 19  & 0.131 &   5.31  & $-0.054 \pm 0.008$ & $1.85 \pm 0.08$ & 17.12 \\
A\,550   & 05 52 51.84 &  $-$21 03 54 &  0.125 &   7.06  & $-0.057 \pm 0.006$ & $2.05 \pm 0.07$ & 17.01 \\
A\,1079 $^{\ddagger}$ & 10 43 24.90 &  $-$07 22 45 &  0.132 & $<$0.45 & $-0.045 \pm 0.003$ & $1.80 \pm 0.05$ & 17.14 \\
A\,1084  & 10 44 30.72 &  $-$07 05 02 &  0.134 &   7.42  & $-0.051 \pm 0.009$ & $1.83 \pm 0.06$ & 17.17 \\
A\,1285  & 11 30 20.64 &  $-$14 34 30 &  0.106 &   5.47  & $-0.056 \pm 0.014$ & $1.87 \pm 0.13$ & 16.61 \\
A\,1437  & 12 00 25.44 &  $+$03 21 04 &  0.133 &   7.72  & $-0.072 \pm 0.005$ & $1.94 \pm 0.12$ & 17.15 \\
A\,1650  & 12 58 41.76 &  $-$01 45 22 &  0.084 &   7.81  & $-0.039 \pm 0.003$ & $1.69 \pm 0.06$ & 16.06 \\
A\,1651  & 12 59 24.00 &  $-$04 11 20 &  0.084 &   8.25  & $-0.047 \pm 0.003$ & $1.63 \pm 0.08$ & 16.06 \\
A\,1664  & 13 03 44.16 &  $-$24 15 22 &  0.127 &   5.36  & $-0.060 \pm 0.015$ & $1.73 \pm 0.13$ & 17.05 \\
A\,2055  & 15 18 41.28 &  $+$06 12 40 &  0.102 &   4.78  & $-0.052 \pm 0.013$ & $1.70 \pm 0.08$ & 16.53 \\
A\,3888  & 22 34 32.88 &  $-$37 43 59 &  0.151 &  14.52  & $-0.097 \pm 0.017$ & $2.12 \pm 0.16$ & 17.47 \\
\hline
\end{tabular}
\begin{tabular}{l}
\small $(\ddagger)$ Although not originally fulfilling the LARCS selection
criteria, A\,1079 resides in the field of A\,1084 at a similar 
redshift to the \\
latter and is thus included in this study.
\end{tabular}
\label{tab:general}
\end{center}
\end{table*}

At higher redshifts, deep multi-colour imaging with {\it Hubble Space
Telescope} ({\it HST}) of distant clusters has sufficient sensitivity
to uncover the signs of the variation in the CMR with environment,
as well as providing detailed morphological information which can be used to
remove the effects of changes in the morphological mix from the analysis
(Ellis et al.\ 1997).  However, the limited field of view of the {\it
WFPC2} camera means that these {\it HST}-based surveys are restricted
to the inner regions of clusters.  For example, van Dokkum et al.\
(1998) discuss the radial dependence of the CMR for the distant cluster
MS\,1358+62 at $z=0.33$ within a $\sim 1.6$\,Mpc region.  They find no
radial dependence of the zero-point of the CMR, or the scatter around it,
for bulge-dominated galaxies at any radius.  They do find, however, 
an increasing scatter around the CMR for early-type disk galaxies in 
the outer parts of their survey region (van Dokkum et al.\ 2000).  They 
propose that this trend is caused by infalling galaxies at the outskirts of
the cluster, but the absence of detailed morphological classifications
for the early-type disk galaxies still leaves open the possibility that
their result may again be due to subtle changes resulting from 
the morphology-density relation.

Spectroscopy provides a much more sensitive tool to investigate the
variation in the ages of the stellar populations of galaxies with
environment, as well as providing unambiguous membership information
for galaxies at large projected distances from the cluster centre.  Spectral
signatures of past activity can remain measurable for $\sim$2 Gyrs
after the star formation has ceased, by which time the photometric
signatures may have become undetectable (e.g.\ Terlevich et al.\ 1999).
Spectroscopic surveys of both local and distant clusters have uncovered
precisely these signatures: identifying a class of galaxies with spectral
features (enhanced Balmer absorption lines) indicative of recently
ended star formation (Couch \& Sharples 1987; Caldwell et al.\ 1993).
The galaxies are classified as  a+k or k+a based on the strength of the
Balmer absorption lines (Dressler et al.\ 1999), equivalent terms are
E+A, PSG and HDS (Couch \& Sharples 1987).  Global gradients in these
spectral indices have been identified in galaxies from a sample of X-ray
selected distant clusters (Balogh et al.\ 1999). However, the samples
available from most spectroscopic studies are relatively modest and for
the most part lack detailed morphological information. Thus there
is a similar ambiguity to that in the analysis of the photometric gradient,
due to the morphology-density relation, and it is difficult to interpret
the origin of the radial gradients observed in key spectroscopic emission
line indicators (Poggianti et al.\ 1999).

One survey which combines the photometric and spectroscopic techniques is
the study of A\,2390 ($z=0.23$) by Abraham et al.\ (1996).  They combine
a large spectroscopic survey covering a wide field (to provide spectral
line indices and membership) and precise photometry.  They demonstrate
that the colours of the reddest spectroscopically-confirmed cluster
members on the colour-magnitude relation become progressively bluer as a
function of clustocentric distance out to 5\,Mpc, a trend which was also
seen in Balmer line indices.  They interpret these radial gradients
as arising from a decrease in the mean luminosity-weighted age of the
stellar populations in galaxies at larger projected distances from the
cluster's centre.  They suggest that this results from the truncation
of star formation as galaxies and groups are accreted from the field
onto the cluster outskirts, building up the cluster in an ``onion-ring''
fashion reminiscent of the hierarchical picture of cluster formation
(Lacey \& Cole, 1993; Cole et al.\ 2000).

The impressive studies of Coma by Terlevich et al.\ (2001) and A\,2390
by Abraham et al.\ (1996) illustrate the power of photometric
comparisons of galaxies across a range of environments to address the
questions posed earlier.  Both studies, however, only cover a single
cluster and so it is not clear how general their conclusions are,
especially given the cluster-to-cluster differences which could arise
from different evolutionary histories.

This paper describes an analysis of the Las Campanas Observatory and
Anglo--Australian Telescope Rich Cluster Survey (LARCS) to address this problem.
LARCS is a long-term project to study a statistically-reliable sample
of 21 of the most luminous X-ray clusters at intermediate redshifts
($z=0.07$--0.16) in the southern hemisphere.  We are mapping the
photometric, spectroscopic and dynamical properties of galaxies in rich
cluster environments at $z\sim 0.1$, tracing the variation in these
properties from the high-density cluster cores out into the surrounding
low-density field beyond the turn-around radius.  For the most massive
clusters at $z\sim 0.1$, the turn-around radius corresponds to roughly
1\,degree or a 10 Mpc radius (O'Hely et al.\ 1998) and 
therefore we have obtained panoramic CCD imaging covering
2-degree diameter fields, as well as spectroscopic coverage of these
fields (Pimbblet et al.\ 2001; O'Hely 2000; O'Hely et al.\ 1998).
The imaging comes from $B$ and $R$-band mosaics taken with the 1-m Swope
telescope at Las Campanas Observatory, while the spectroscopy comes from
the subsequent follow-up with the 400-fibre 2dF multi-object spectrograph
on the 3.9-m Anglo-Australian Telescope (AAT).

The advantage of this sample for the present analysis is the
homogeneity of the clusters and observations.  This allows us to combine many
clusters together to provide a large sample of cluster galaxies across
a wide range in environment.  In particular, we are able to use
statistical corrections to remove field contamination, rather than
requiring spectroscopic membership.  
We can thus investigate the colour of galaxies
across a wide range in environments in clusters, spanning three orders
of magnitude in projected galaxy density from the core out to
close to the turn-around radius at $\sim 8$ Mpc.

In \S2, we describe the LARCS cluster sample and the precision of our
photometric catalogues.  In \S3 we outline the statistical method used
to correct for field contamination and the construction of colour
magnitude diagrams.  In \S4, we present our results on
the characteristics of the
galaxy populations in the clusters.  By creating a composite cluster
we explore the variation in the CMR with radius and local galaxy density.
We discuss these results in \S5 and present our conclusions in \S6.

%
%
\begin{figure*} 
\begin{centering} 
\psfig{file=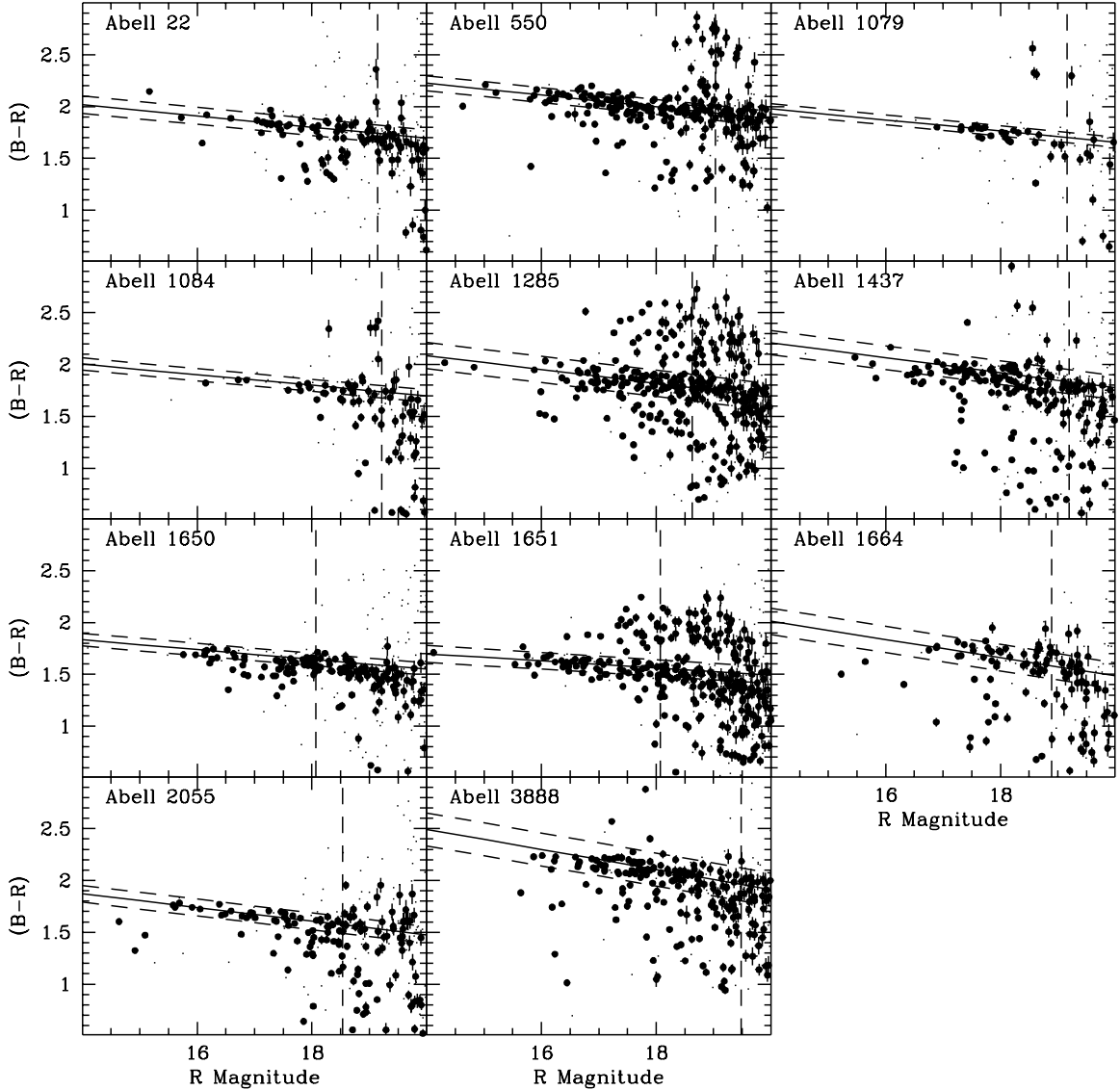,width=17cm}
\caption{\small{
The colour-magnitude diagrams for the 11 clusters used in this
work.  All galaxies statistically-identified as cluster members 
within a 2\,Mpc radius of the cluster centre from one
realization of the field subtraction are plotted as heavy points.  
The smaller points are the galaxies which were rejected
as field by our statistical subtraction method.  The biweighted
fit to the CMR is shown as a solid line.  The 1$\sigma$ uncertainty 
in the colour is shown by the flanking parallel dashed lines.   
The vertical line denotes the absolute magnitude limit
of $M_V=-20$ used to define the galaxy sample used for the fit.
}}
\label{fig:inners}
\end{centering}
\end{figure*}

%
%
\section{Cluster selection and observations}

The sample used in our analysis is from the LARCS survey (Pimbblet et
al.\ 2001). LARCS is a survey of the properties of an X-ray
luminosity-limited sample of 21 clusters at $z=0.07$--0.16.  The
clusters in the sample are selected from the X-ray brightest
Abell Clusters (XBACS) catalogue of Ebeling et al.\ (1996).  XBACS
comprises 242 Abell clusters detected in the {\it ROSAT} All-Sky Survey
and is effectively a complete, volume-limited sample of the most X-ray
luminous clusters, $L_X \gs 5 \times 10^{44}\ergs$, at $z \sim 0.1$.
The 21 clusters in the LARCS sample are a random subsample of the
southern XBACS clusters with $L_X > 3.7 \times 10^{44}\ergs$ at
$z=0.07$--0.16.

Ten clusters from LARCS are analysed in this paper and we list them in
Table~\ref{tab:general} (an additional cluster, A\,1079, is present in
our observations of A\,1084 and we have included it in our analysis).
These ten clusters possess the highest quality and most complete
photometric observations from the full survey.  The subset covers a
representative range of the complete LARCS sample, including clusters
which appear relaxed in their X-ray and galaxy distributions
(e.g.\ A\,1650), as well as more dynamically young, disturbed looking
systems (e.g.\ A\,1664 -- which appears to have undergone a recent
cluster--cluster merger; Edge, priv. comm.).

High quality broad-band $B$- and $R$-band CCD images of each cluster
have been obtained at Las Campanas Observatory using the 1-m Swope
Telescope.  More details of the observations, reduction and analysis of
these data are given by Pimbblet et al.\ (2001); here we briefly
summarise the pertinent points.

A $2\deg$-diameter field around each cluster is imaged in 21
over-lapping pointings to produce a mosaic of a region out to $\sim
10$\,Mpc radius at the survey's median redshift of $z\sim0.12$.  The
images are reduced using standard tasks within {\sc iraf} and are then
catalogued using the SExtractor package of Bertin \& Arnouts (1995).  
We adopt ``{\sc mag\_best}'' from SExtractor as the estimate of the 
total magnitudes.  
To determine colours for the sources we measure aperture photometry
within $4''$ diameter apertures (typically $\sim10$\,kpc at the cluster
redshifts) on seeing-matched tiles using {\sc phot} within {\sc iraf}.
Photometric zeropoints are computed using frequent observations of
standard stars (typically over 100 standard stars per broad-band filter
per night) from Landolt (1992) interspersed throughout the science
observations.  On photometric nights, the variation in the resultant
colour and extinction terms are all within $1\sigma$ of each other and
the final photometric accuracy is better than 0.03 mags.

As some of the observations within a mosaic are undertaken in
non-photometric conditions it is necessary to calibrate these
independently.  This is achieved through the method developed by
Glazebrook et al.\ (1994).  We use information on the photometry of
objects in the overlapping regions of the tiles to calculate the
offsets between photometric and non-photometric mosaic tiles.  Once
corrected by Glazebrook's algorithm, the resultant median tile-to-tile
variation in zero-points across the mosaic is $< 0.006$\,mags (see
Figure~1 of Pimbblet et al.\ 2001), as estimated from the scatter in
the magnitude from the duplicated sources in the overlapping
regions.  The maximum deviations between tiles is $\sim 0.015$\,mags.
We expect that the photometric zeropoint errors of 0.03\,mags dominate
the systematic error budget and the tile-to-tile variation in
zero-points are small by comparison.

Finally, we apply corrections for galactic reddening based on
Schlegel et al.\ (1998).  The final internal magnitude errors across
the full mosaics are typically 0.03 mag and always less than
0.06 mags (Pimbblet et al.\ 2001).  Such accuracy is essential to
detect subtle photometric radial gradients in the colour-magnitude
relation.  The catalogues are typically 80 percent complete at a depth
of $R\sim22.0$ and $B\sim23.0$.

Star/galaxy separation for the catalogues uses the robust criteria
described in Pimbblet et al.\ (2001): FWHM\,$ > 2.0''$ and {\sc
class\_star}\,$ < 0.1$ from the SExtractor star--galaxy classifier.
The resulting stellar contamination based on these criteria is
estimated at $\leq 3$ percent (see Pimbblet et al.\ 2001).  To further
check for misclassification of compact early-type cluster galaxies as
stars we have constructed colour-magnitude diagrams for the star
samples and confirm that they exhibit no hint of a CMR at the relevant
colour for the cluster members.

%
%

\section{Construction of the Colour-Magnitude diagrams}

In this section we describe the construction of colour-magnitude
diagrams from the LARCS catalogues, the use of the biweight method to
fit the CMR and the spatial distribution of the cluster galaxies (i.e.\  cluster
morphology).

\subsection{Statistical field correction}

To examine the galaxy population of the clusters in the absence of
spectroscopy it is necessary to correct for the ``field''
contamination in a given cluster catalogue.  The statistical subtraction
method utilized in this study is described in detail in Appendix~A.

Briefly, the field population is determined from the outer regions of
the LARCS cluster mosaics at a radius well beyond 6\,Mpc (at a redshift
of $z=0.12$).  The field regions from each cluster are all examined by eye
prior to being stacked to give the average field distribution.  
After examining these data, we reject part of the field around
Abell~1084 due to the presence of Abell~1079.
The field sample is not significantly biased to a lower 
mean density because of its removal.  
Once generated, the final
stacked field population\footnote{The
final effective area of our field sample is $\sim13.1$ square degrees.
Each cluster mosaic covers $\sim3.1$ square degrees on the sky.}
is then appropriately area-scaled
to that of the cluster sample we wish to correct.

The colour-magnitude diagrams for galaxies from the cluster sample and
the final, scaled, stacked field sample are then compared.  By direct
comparison of corresponding regions on the colour-magnitude diagrams
we assign each galaxy in the cluster sample a probability of being a
field galaxy.  Then, using a Monte Carlo method, we subtract off the
field population based upon these probabilities.  As described in
Appendix~A, however, problems can arise if the calculated probability 
exceeds 1.0. The solution utilized in this work is to expand the 
interval in colour and magnitude used to calculate the probability
until it lies in the range $0.0<$ P(Field) $<1.0$ 
(see Appendix~A for a full discussion).

In this work, the statistical Monte Carlo background subtractions are
realized 100 times and the colour-magnitude fits presented  
represent the median of those background subtractions.

\subsection{Fitting the CMR}

We run the background subtraction algorithm (see Appendix~A) on each
cluster to produce 100 realizations of the cluster's colour-magnitude
diagram.  In fitting the CMR we only use those galaxies whose absolute
rest frame magnitude is brighter than $M_V = -20$.  This magnitude is
chosen to correspond to the equivalent limit in Butcher \& Oemler
(1984).  

Each of the CMRs are fitted using a robust biweight method (Beers et
al.\ 1990) identical to that employed by Terlevich (1998).  Briefly,
the biweight method is a statistically robust estimator that weights
outlying points (such as the blue population) low in order to find the
best fit line to the CMR.  In line with Beers et al.\ (1990), we adopt
the median absolute deviation as the best estimator of the biweight
location and scale and a value of the weighting parameter of $w=6$, this
results in zero-weight being given to galaxies more than 4$\sigma$ away
from the median relation with the weighting increasing quadratically
closer to the median value (e.g. 95 per cent weighting at $1\sigma$).  
Since we seek to fit a model to these data,
we must minimize the residuals of the data to the model.  Therefore the
set of points $x_i = Y_i - f(X_i)$ are defined, where $(X_i,Y_i)$ are
the data points and $f(X)$ is the model (a straight line of the form
$f(X)=mX+c$).  The scale and location of the $x_{i}$ are then
iteratively minimized on the $(m,c)$ plane using the multidimensional
downhill simplex method of Press et al.\ (1992).  We apply this routine
to the background-subtracted cluster sample from each of the 100 Monte
Carlo simulations.  We adopt the median of the fit parameters from the
100 simulations as our best estimate of the CMR and use the scatter
between the simulations to give the uncertainty in this fit.  We note
that our adoption of the errors from the Monte Carlo field correction
may underestimate the uncertainty in the fits when the sample is
dominated by the cluster population.  Most of our analysis, however, lies
in the regime where the field contribution is important and so we
retain these errors as the best estimates of the internal
uncertainty in our fits.

In Figure~\ref{fig:inners} we show the distribution of galaxies on
colour-magnitude diagrams for one such realization of each of the
clusters within a radial extent of $r_p<2$\,Mpc.  Details of the median
values of the fits to the 100 realisations for each cluster and their
associated errors are presented in Table~\ref{tab:general}.

\begin{figure} 
\begin{centering}
\psfig{file=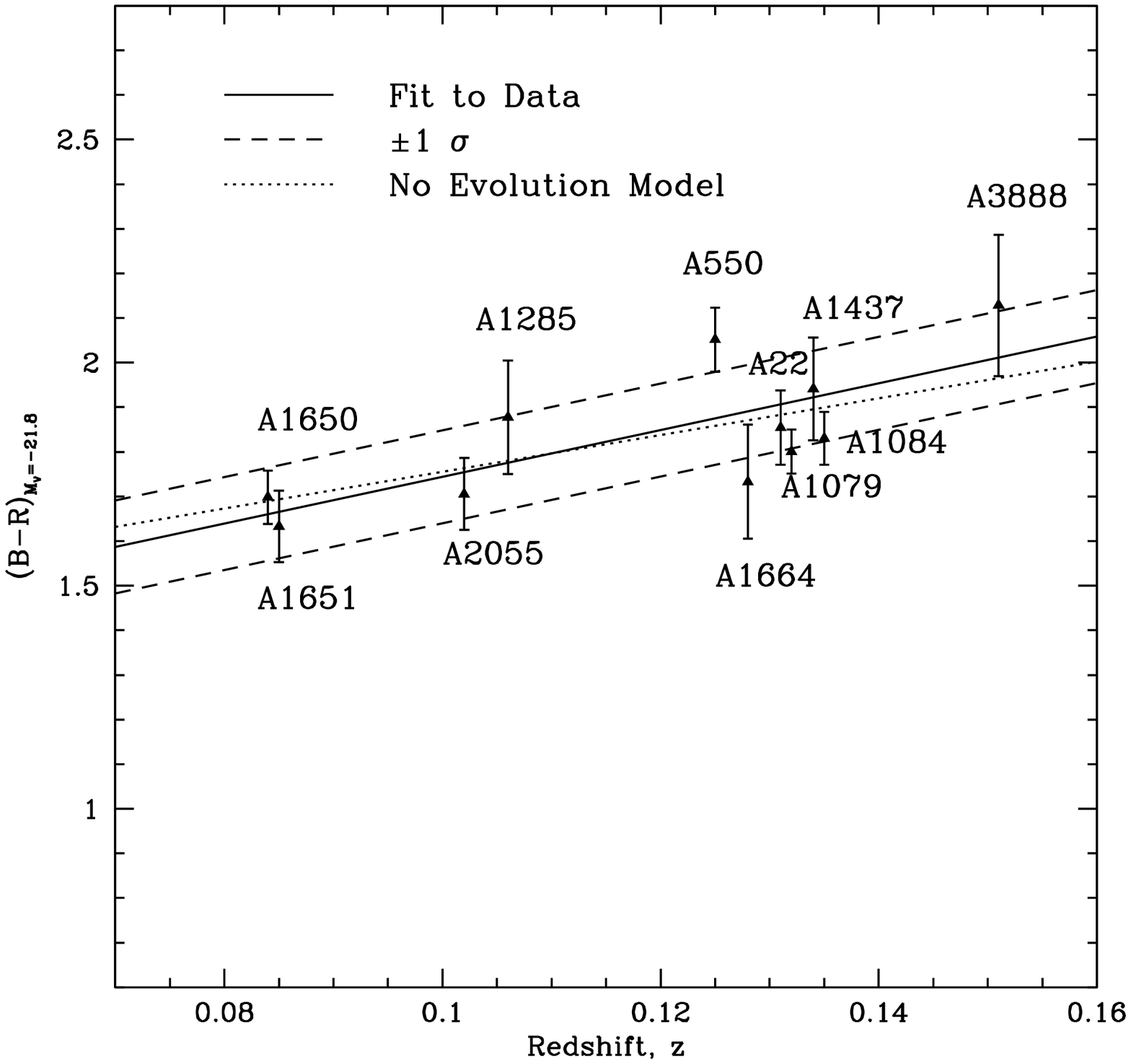,width=3.in,height=3.in} 
\smallskip
\psfig{file=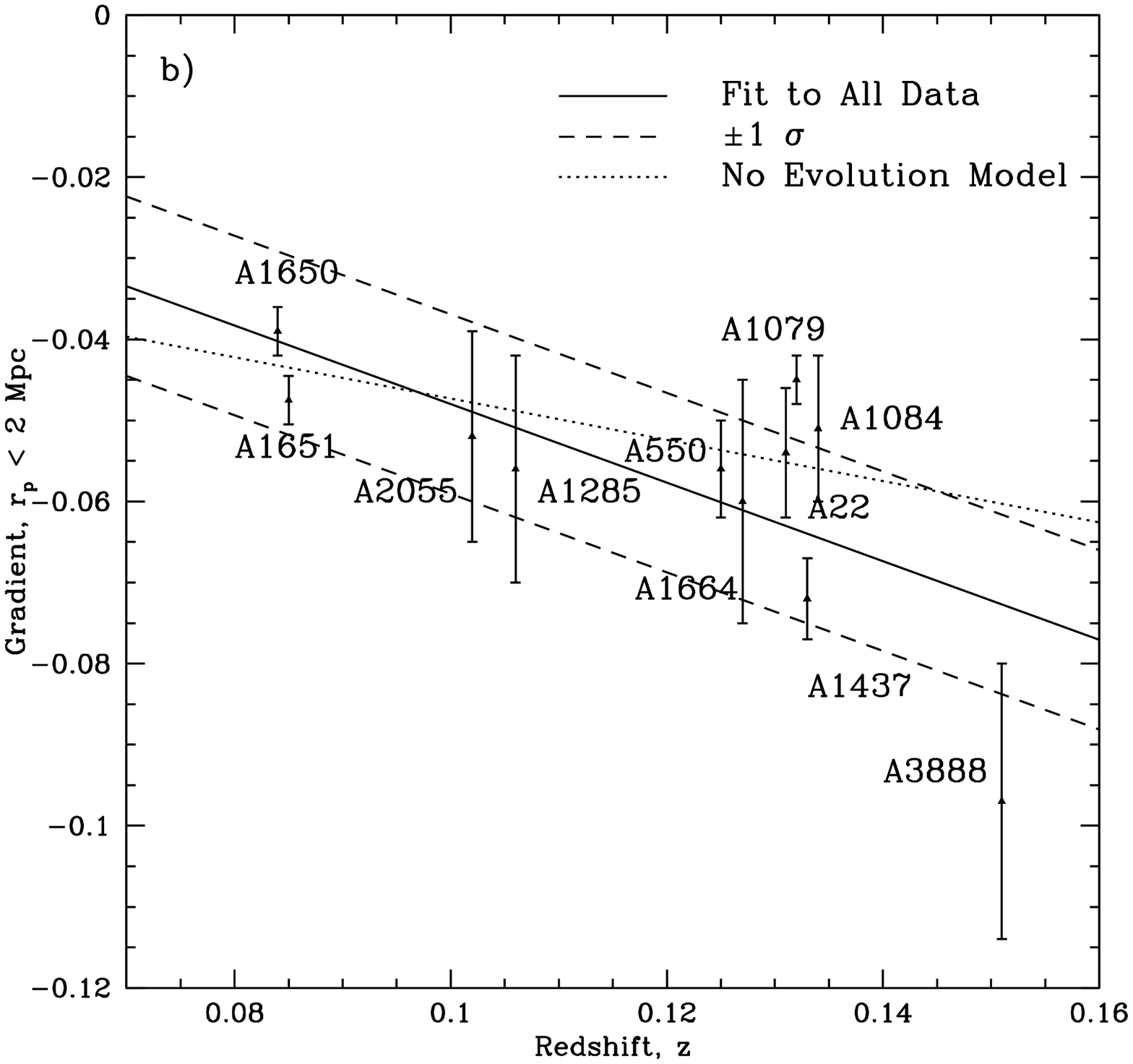,width=3.in,height=3.in} 
\caption{\small{{\bf a)}  Median observed $(B-R)$ colour at
$M_V=-21.8$ calculated from 100 realizations of the background subtraction
plotted versus cluster redshift (with no K-correction applied).  The solid
line is the best fit to the data, with the parallel dashed lines being
$\pm 1 \sigma$.  The dotted line illustrates the expected evolution
in the galaxy colours of galaxies which formed all their stars at high
redshift, $z\gg 2$.  The clusters form a tight relation evolving redward
with increasing redshift.   ~~~ {\bf b)}  The variation
in slope of the CMR with redshift.  There is a trend for steeper
CMR's at higher redshifts, as expected from the effects of shifting
K-corrections on the observed $(B-R)$ colours of the galaxies along
the CMR.  We show a ``No evolution'' model which shows the behaviour
expected just from the differential K-corrections.
}}
\label{fig:ZvsC}
\end{centering}
\end{figure}

All the clusters show colour-magnitude relations for the redder,
early-type galaxy members.  We see strong CMRs 
(more than 60 galaxies lying on the CMR within 2 Mpc of the cluster
centre)
in A\,550, A\,1285 and A\,3888; whilst A\,1079, A\,1084 and
A\,1664 exhibit much weaker relations (less than 25 galaxies lie on 
the CMR within 2 Mpc of the cluster centre).  
All the CMR's exhibit a
negative slope as expected from the colour-luminosity relation of
early-type galaxies (Table~\ref{tab:general}, Figure~\ref{fig:ZvsC}).

The colour at a fiducial magnitude of $M_V=-21.8$ (equivalent to an
$L^\star$ galaxy) is also calculated from the result of the biweight
fit; $(B-R)_{M_V=-21.8}$.  These colours are presented in
Figure~\ref{fig:ZvsC} and tabulated along with the values of
$M_V=-21.8$ in Table~\ref{tab:general}.  The $(B-R)_{M_V=-21.8}$
colours of galaxies on the CMRs in the clusters form a tight relation,
median deviation of 0.08\,mags, which steadily reddens with redshift
due to the K-correction.  We also
show the expected variation with redshift in the observed colour of a galaxy
which formed all its stars at very high redshift, $z \gg 2$.  The
reddening in colour at the fiducial magnitude we see is broadly in
line with that expected for a population which formed the bulk of its
stars at high redshift (Metcalfe et al.\ 1991).
The small scatter in the redshift trend 
implies that our cluster to cluster photometry is internally consistent.

%
%

The statistical field correction we apply is designed to detect
overdensities above the level of the ``average'' field on the
colour-magnitude plane.  Therefore, if there are other over-dense
regions in the fields of our clusters we expect to see some secondary
CMRs in the colour-magnitude diagrams.  Such CMRs are seen in the
colour-magnitude diagrams of A\,550, A\,1285 and A\,1651.  As expected
from the available volume, these CMRs are higher redshift clusters in
the background of the observed clusters, for example, we identify the
background cluster in A\,1651 ($z=0.084$) as A\,1658 ($z \sim 0.145$).
To reduce the influence of these systems on the biweight estimator used
to fit the CMR of the LARCS cluster, we employ an appropriate red
limit to ensure the fit converges on the desired CMR.

\subsection{The spatial distribution of cluster galaxies}

\begin{figure*} 
\begin{centering} \vspace*{-2cm}
\psfig{file=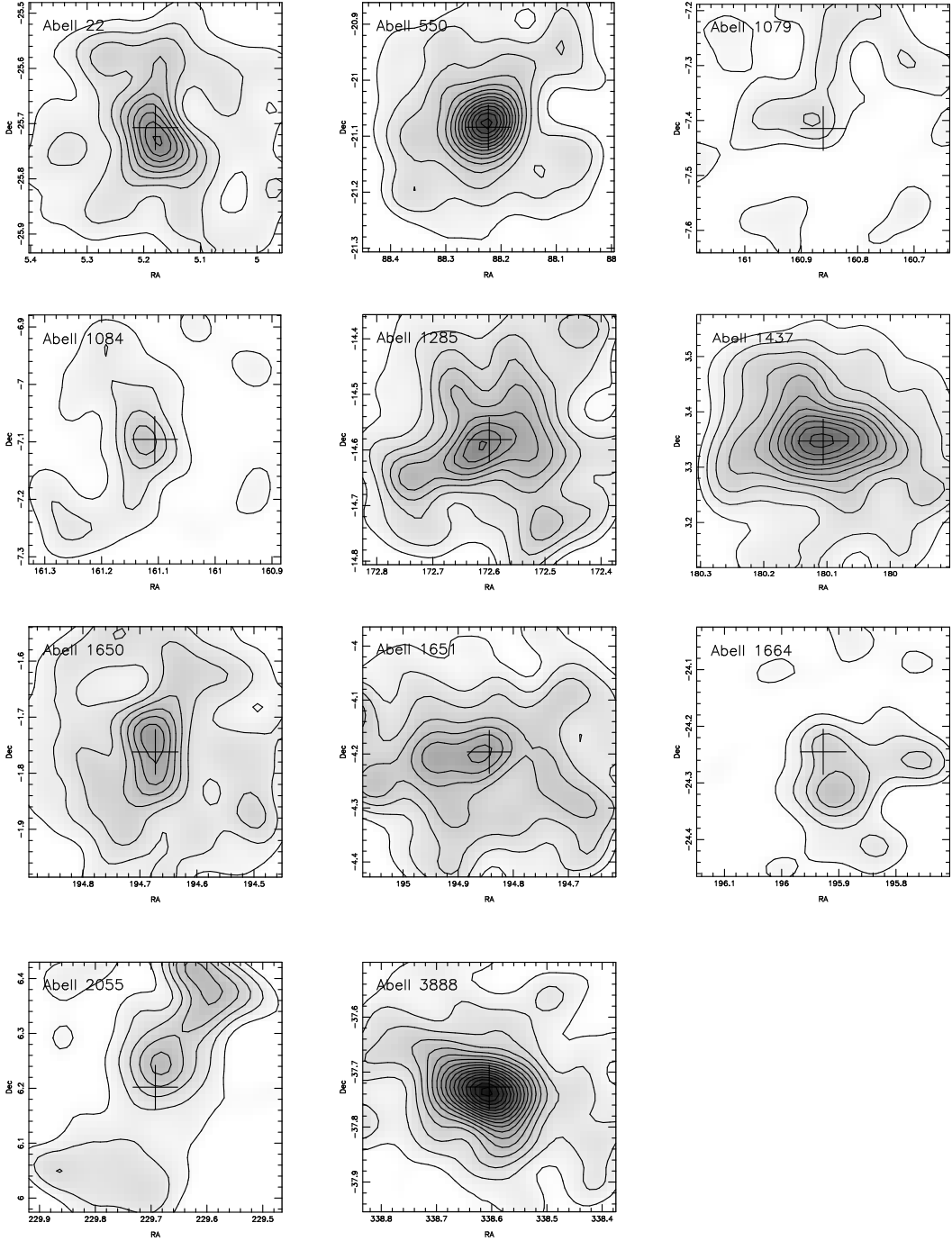,width=17cm} \vspace*{-1.5cm}
\caption{\small{The two dimensional smoothed spatial distributions of the
cluster galaxies lying within the scatter of the biweight fit to the
CMR.  A circular top-hat function with a 500\,kpc smoothing length 
is used on the positions of
galaxies to generate these maps.  The lowest contour represents a
surface density of cluster galaxies of 6.0\,Mpc$^{-2}$, with each
subsequent inward contour denoting an increase of 2.0\,Mpc$^{-2}$.  The
crosses indicate the adopted centre of the clusters based on the
position of the apparent brightest cluster member from the CMR.
}}
\label{fig:maps}
\end{centering}
\end{figure*}

We now use the CMRs to map the two dimensional galaxy distribution
within the clusters.  Assuming that the galaxies on the CMR are
associated with the cluster, we define a cluster galaxy to be within
the $1\sigma$ scatter of its fitted colour-magnitude relation  (listed
in Table~\ref{tab:general}).  This criteria is applied to the whole sample
of galaxies in each field.  Having obtained their positions on the sky,
we use a circular top-hat function with a 500\,kpc smoothing 
length to generate smoothed galaxy distribution
maps for each cluster.  The resultant distributions for the central $25'$
of each cluster are presented in Figure~\ref{fig:maps}.

The brightest galaxy in these X-ray luminous clusters is typically a giant
elliptical
and we identify these using the CMR and the morphology of their
extended low surface brightness halo in our imaging data.  For the more
regular clusters, the position of this galaxy is generally in agreement
with the peak of the distribution of colour-selected cluster members
(Figure~\ref{fig:maps}) and we therefore use the position of this
galaxy to define the cluster centre.  Our clusters demonstrate a broad
range of morphological types ranging from very regular,
highly-concentrated examples (e.g.\ A\,550), through to very diffuse
systems (e.g.\ A\,1664) whose central galaxy position does not coincide
with any of the main spatial overdensities seen in the colour-selected
cluster members.

%
%

%
%

%
%
\begin{figure*} 
\begin{centering}
\psfig{file=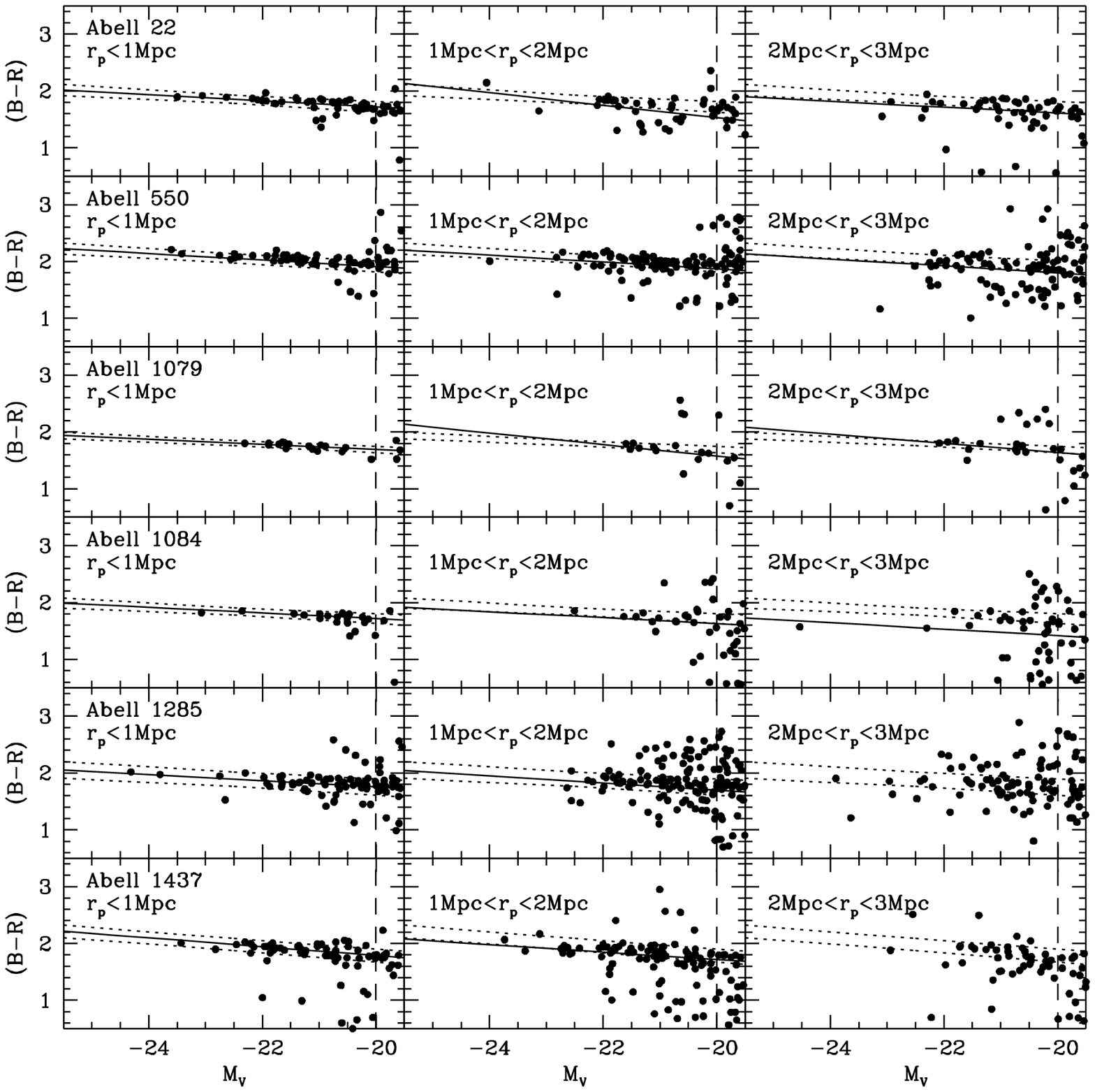,width=17cm} 
\caption{(a)\small{
Colour-magnitude diagrams for A\,22, A\,550, A\,1079, A\,1084, A\,1285
and A\,1437 in 1-Mpc wide radial bins.  Each diagram shows one of the 100
realizations made with the background subtraction algorithm.  The solid
line, where present, is the biweighted fit to the CMR; 
if absent, a biweighted fit could not be successfully made.
The short dashed lines denote the combined 
errors associated with the $r_p<1$\,Mpc fit of each cluster.
The longer dashed, vertical lines denote the $M_V=-20$ fiducial magnitude 
limit; only galaxies brighter than this are considered in the biweighted fit.
For most of the clusters, there is a trend for the CMR to evolve bluewards
by $\Delta (B-R) \sim -0.09\pm 0.11$ out to 3\,Mpc.
}}
\label{fig:cmds1}
\end{centering}
\end{figure*}

\section{Analysis and Results}

In this section we describe the analysis of the properties of the
cluster's CMRs as a function of radius and local galaxy density.

\subsection{Radial variation of the CMR}

%
%
\setcounter{figure}{3}
\begin{figure*} 
\begin{centering}
\psfig{file=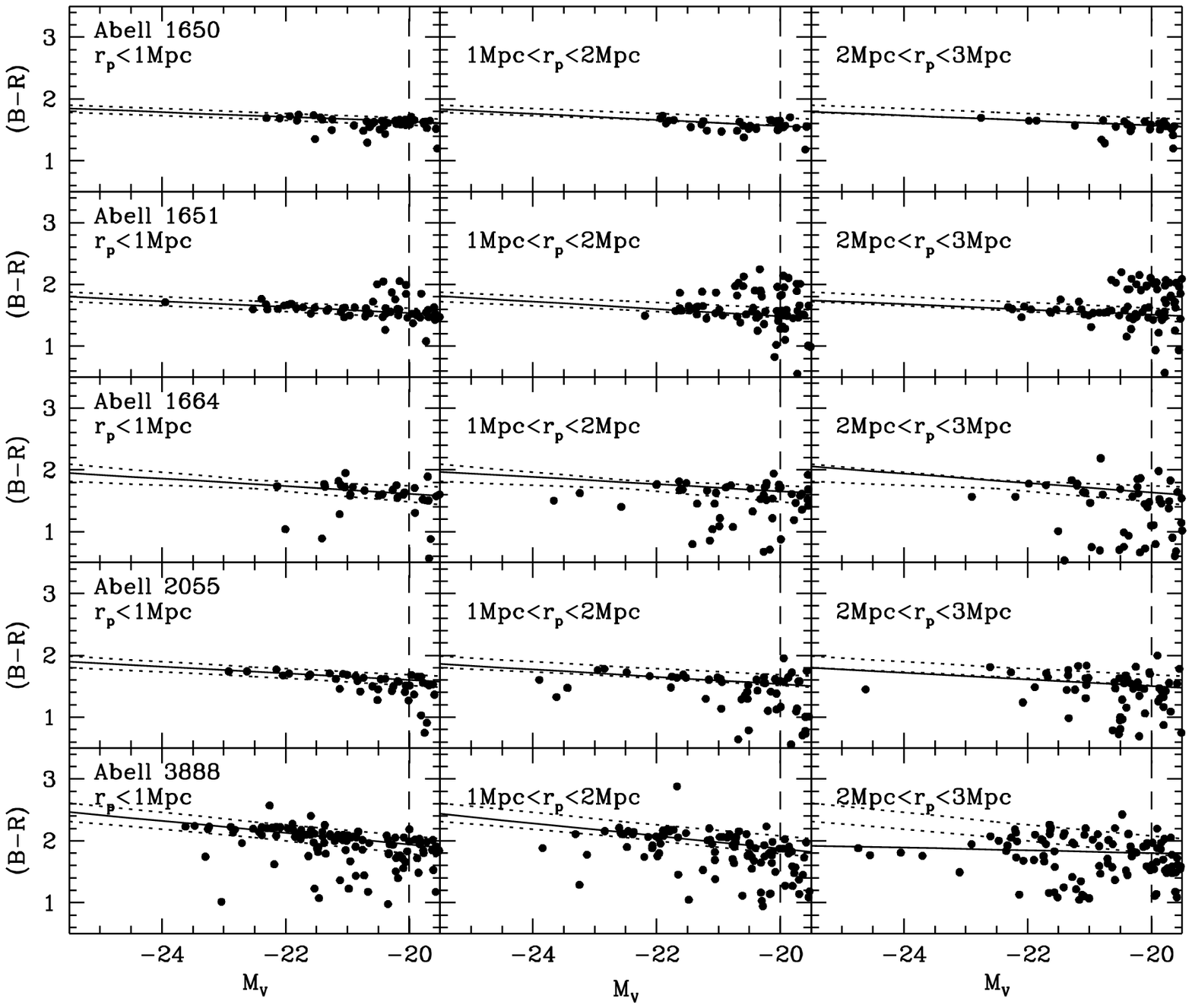,width=17cm}
\vspace{-1.in} 
\caption{(b)\small{
As for Figure~\ref{fig:cmds1}a, but with A\,1650, A\,1651, A\,1664,
A\,2055 and A\,3888.
}}
\label{fig:cmds2}
\end{centering}
\end{figure*}

We search for any radial variations in the CMR within the clusters by
dividing the central 6\,Mpc diameter region of clusters into three 
anulii.  We note that the virial radii for these X-ray luminous
clusters should lie in the range 2--4 Mpc (Carlberg et al.\ 1996).  For each
radial bin, a biweighted fit is calculated to the CMR and the colour at
a fiducial magnitude (equivalent to $M_V=-21.8$) is obtained.  We have
also calculated the colour based on a fixed slope  (using the fit to
the CMR in the central 2\,Mpc diameter annulus) for the
field-subtracted cluster realizations and found that this method
produces the same results as an unconstrained gradient fit, within 1 standard
deviation (see Table~\ref{tab:fits2}).  The result of these fits are
presented in Table~\ref{tab:fits2}, whilst the colour-magnitude
relations for the three radial bins in each cluster can be
seen in Figure~4.

In most of our clusters we observe a modest blueward shift 
in colour at the fiducial magnitude with radius by $\Delta (B-R) 
\sim -0.09\pm 0.11$ out to 3\,Mpc (Table~\ref{tab:fits2}).
We see no significant change in this trend with cluster redshift.
Discrepant results arise in three clusters.  In A\,1084, the change in
$(B-R)$ colour from the central radial bin to the third is much more
than $\Delta(B-R) \sim -0.10$.  We attribute this to the poor fit in the
third radial bin (see Figure~\ref{fig:cmds1}).  The third radial bin of
A\,3888 is similarly poorly fitted.  In A\,1664, no radial trend is
apparent.

However, we note that there are two shortcomings of this analysis.
Firstly, the formal errors on the biweighted fits for individual
clusters are large enough that the colour gradient trend in the individual CMRs
is insignificant.  Secondly, the biases in the statistical correction
of the colour-magnitude diagram make it increasingly unreliable at
large radii (Appendix~A).  We must therefore adopt a different approach.

%
%
%
\begin{table*}
\begin{center}   
\caption{\small{Parameters of the biweight fit to the CMRs of the clusters, split into
three equal radial bins as plotted in Figure~4.  
The slope and colours are those found using the
biweight estimator fit to each radial bin.  The values reported in the
final column (fixed slope) uses the slope of the central bin in the
particular cluster.
The errors are the $1\sigma$ variation from the 100 Monte Carlo 
simulations of the background subtraction.
}}
\begin{tabular}{lcccc}
\hline
Cluster    & Radius  &	Slope	&	$(B-R)_{M_V=-21.8}$ 	& $(B-R)_{M_V=-21.8}$	\\
	   & (Mpc)		  &		&			        & Fixed Slope \\
\hline \hline
A\,22        & 0--1 & $-0.054 \pm 0.008$ &	$1.85 \pm 0.07$ 	& $1.85 \pm 0.07$ \\ 
	     & 1--2 & $-0.110 \pm 0.010$ &	$1.87 \pm 0.12$ 	& $1.73 \pm 0.21$ \\
	     & 2--3 & $-0.052 \pm 0.007$ &	$1.75 \pm 0.17$  	& $1.73 \pm 0.22$ \\ 
\noalign{\smallskip}
A\,550       & 0--1 & $-0.057 \pm 0.001$ &	$2.05 \pm 0.07$  	& $2.05 \pm 0.07$ \\
	     & 1--2 & $-0.060 \pm 0.003$ &	$2.02 \pm 0.15$ 	& $2.02 \pm 0.14$ \\
	     & 2--3 & $-0.061 \pm 0.014$ &	$1.94 \pm 0.35$ 	& $1.92 \pm 0.38$ \\
\noalign{\smallskip}
A\,1079      & 0--1 & $-0.045 \pm 0.003$ &	$1.80 \pm 0.05$ 	& $1.80 \pm 0.05$ \\ 
	     & 1--2 & $-0.101 \pm 0.004$ & 	$1.82 \pm 0.07$ 	& $1.76 \pm 0.09$ \\
	     & 2--3 & $-0.081 \pm 0.009$ & 	$1.82 \pm 0.20$ 	& $1.85 \pm 0.33$ \\
\noalign{\smallskip}
A\,1084      & 0--1 & $-0.050 \pm 0.009$ & 	$1.83 \pm 0.06$  	& $1.83 \pm 0.06$ \\
	     & 1--2 & $-0.052 \pm 0.014$ & 	$1.78 \pm 0.42$ 	& $1.79 \pm 0.31$ \\
	     & 2--3 & $-0.057 \pm 0.027$ & 	$1.55 \pm 0.54$ 	& $1.66 \pm 0.56$ \\
\noalign{\smallskip}
A\,1285      & 0--1 & $-0.055 \pm 0.007$ & 	$1.87 \pm 0.12$ 	& $1.87 \pm 0.14$ \\
	     & 1--2 & $-0.060 \pm 0.013$ & 	$1.86 \pm 0.33$ 	& $1.86 \pm 0.34$ \\
	     & 2--3 & ... 			& ...  			& ...		   \\
\noalign{\smallskip}
A\,1437      & 0--1 & $-0.077 \pm 0.005$ & 	$1.94 \pm 0.10$ 	& $1.94 \pm 0.12$ \\
	     & 1--2 & $-0.065 \pm 0.003$ & 	$1.87 \pm 0.21$ 	& $1.87 \pm 0.21$ \\
	     & 2--3 & ... 			& ...  			& ...		   \\ 
\noalign{\smallskip}
A\,1650      & 0--1 & $-0.038 \pm 0.001$ & 	$1.70 \pm 0.06$  	& $1.70 \pm 0.06$ \\
	     & 1--2 & $-0.050 \pm 0.005$ & 	$1.65 \pm 0.08$ 	& $1.65 \pm 0.08$ \\
	     & 2--3 & $-0.041 \pm 0.011$ & 	$1.63 \pm 0.07$ 	& $1.65 \pm 0.06$ \\ 
\noalign{\smallskip}
A\,1651      & 0--1 & $-0.047 \pm 0.002$ & 	$1.63 \pm 0.07$  	& $1.63 \pm 0.08$ \\
	     & 1--2 & $-0.060 \pm 0.011$ & 	$1.63 \pm 0.16$ 	& $1.64 \pm 0.24$ \\
  	     & 2--3 & $-0.048 \pm 0.008$ & 	$1.59 \pm 0.15$ 	& $1.60 \pm 0.14$ \\ 
\noalign{\smallskip}
A\,1664      & 0--1 & $-0.062 \pm 0.012$ & 	$1.73 \pm 0.07$  	& $1.73 \pm 0.07$ \\
	     & 1--2 & $-0.058 \pm 0.039$ & 	$1.76 \pm 0.07$ 	& $1.76 \pm 0.08$ \\
	     & 2--3 & $-0.076 \pm 0.024$ & 	$1.76 \pm 0.08$ 	& $1.76 \pm 0.08$ \\ 
\noalign{\smallskip}
A\,2055      & 0--1 & $-0.054 \pm 0.007$ & 	$1.71 \pm 0.08$  	& $1.71 \pm 0.08$ \\
	     & 1--2 & $-0.050 \pm 0.021$ & 	$1.67 \pm 0.15$ 	& $1.62 \pm 0.18$ \\
	     & 2--3 & $-0.054 \pm 0.024$ & 	$1.67 \pm 0.26$ 	& $1.60 \pm 0.28$ \\
\noalign{\smallskip}
A\,3888      & 0--1 & $-0.096 \pm 0.015$ & 	$2.13 \pm 0.13$  	& $2.12 \pm 0.13$ \\
	     & 1--2 & $-0.103 \pm 0.043$ & 	$2.08 \pm 0.22$ 	& $2.04 \pm 0.25$ \\
	     & 2--3 & $-0.193 \pm 0.013$ & 	$1.88 \pm 0.35$  	& $1.90 \pm 0.36$ \\
\hline
\end{tabular}
  \label{tab:fits2}
\end{center}
\end{table*}

To trace the variation in the colour of the CMR out to larger radii
and better quantify its strength, we can exploit the homogeneity of
our cluster sample and observations by combining all of our clusters
together into a single composite system.  This will improve the statistics for
cluster members at large radii and provide a more robust measurement of
the trends in the CMR.  In doing so, however, we acknowledge that
there may be aperture bias present in our photometry: we use a fixed
photometric aperture and therefore cluster galaxies will be sampled 
to different physical radii as a function of redshift.  This effect
may alter our determined CMR slopes and colour changes if there
are strong colour gradients present within our cluster galaxies.
We estimate that at least $85$ per cent of the total light 
falls into our large $4''$ apertures ($\sim 10$ kpc at $z\sim 
0.12$; see \S2) in all but the largest of our galaxies.  
As the nature of this work is a differential radial analysis 
of the composite CMR, we consider such an effect to be small.

To produce a composite cluster we have to transform the magnitudes,
colours and positions of the galaxies in the individual clusters
onto a common scale.  
The transformation of the magnitudes uses the
relative magnitudes of non-evolving early-type $L^\star$ galaxies
from Table~\ref{tab:general}, an assumption which is supported by the
agreement between the no evolution predictions for colour evolution and
the observations across the limited redshift range spanned by our sample.
The apparent magnitudes of the galaxies can therefore be transformed to
a median redshift of $z=0.12$.  To transform the colours of the galaxies
on the CMR onto a uniform scale we simply reduce the observed CMR's to
provide colours relative to the fitted relation.  These can then be easily
transformed to $z=0.12$.  The slope we adopt for this transformation is
assumed to be constant with radius, and equivalent to the measured values
presented in Table~\ref{tab:general}.  
We emphasize that our slopes are determined in a totally homogeneous
manner for a homogeneous sample: two absolute requirements
given the uncertainties.
The most meaningful method to compare
the radial positions of galaxies in the various clusters is to use their
radius normalised to the virial radius of the clusters, $R_{\rm vir}$.
However, the weak dependence of the virial radius on the X-ray 
luminosity, $R_{\rm vir} \propto L^{1/6}$ (Babul et al.\ 2001), 
combined with a narrow X-ray luminosity range for our sample, means 
that $R_{\rm vir}$
is only expected to vary by 20 percent across the whole sample.  
Hence, for simplicity, the clusters are all scaled to a fixed metric
size.

The final step in our analysis of the composite cluster is 
to construct colour histograms down to the fiducial magnitude limit, 
$M_V=-20$.  These histograms are statistically corrected using the 
binned colour distribution from the field.  
We can use this approach as we are interested in the typical
colour of the CMR, rather than its slope, so we no longer need to retain
the magnitude information on the galaxies.  The advantage of using colour
histograms is that it circumvents some of the concerns about the field
correction techniques used on the colour-magnitude plane (Appendix~A).

%
%
\begin{figure*} 
\begin{centering}
\psfig{file=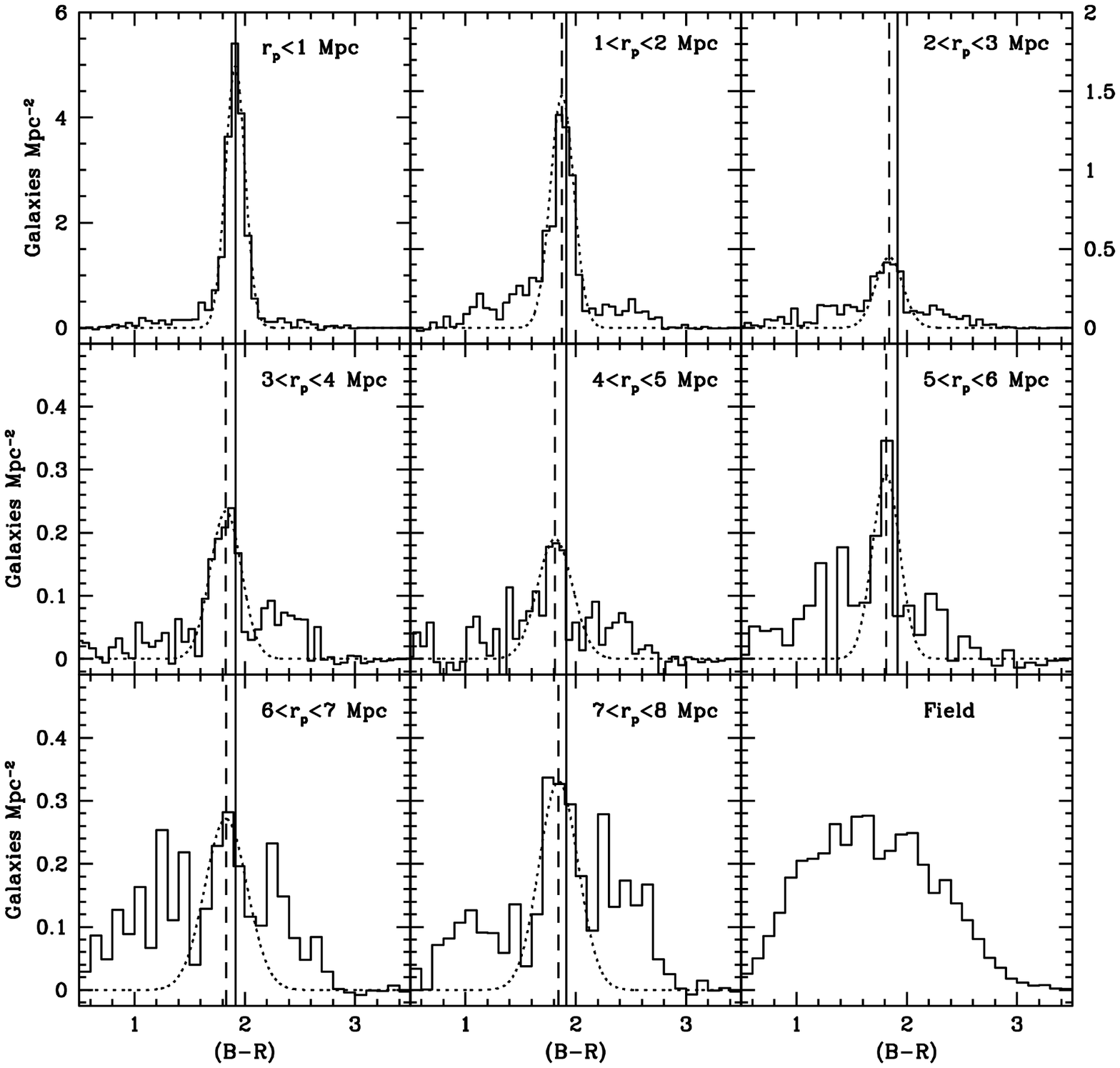,width=17cm} 
\caption{\small{Colour histograms of the composite cluster from the
center to a maximum radial extent of 8\,Mpc.  The field colour
distribution is also pictured in the lower-right for reference
(arbitrarily scaled).  The best fit gaussians are shown for the CMR
peaks in each radial bin.  The solid vertical line is the colour of the
CMR peak in the central bin.  The dashed vertical line is the peak of
each individual radial bin.  A blueward shift in the peak of the CMR
can be seen clearly even by the second radial interval, although the
position of the CMR peak becomes more uncertain at large radii.  The
CMR evolves blueward at a rate of $d(B-R) / d r_p = -0.022 \pm 0.004$ 
from the centre to the outer regions.  Also note that the CMR appears 
to the broaden at larger radius. 
}}
\label{fig:combhists}
\end{centering}
\end{figure*}

We create the colour histogram of the field sample in an identical manner
to the cluster sample.  An appropriately area-scaled region of the field
is subtracted off the composite cluster histogram in each radial bin.
The result of this analysis is illustrated in Figure~\ref{fig:combhists}.

Figure~\ref{fig:combhists} shows that the red cluster members in the CMR
are still visible as a peak in the colour histogram as far out as $\sim
6$\,Mpc from the cluster core.  Beyond this, the red galaxies become
swamped by an increasingly blue cluster population and the noise from
the statistical field correction.  To identify the peak in the CMR and
quantify how it changes with radius we fit the colour histograms with
a gaussian using a $\chi^2$ minimization method.  
To make the fit we restrict the colour range to $(B-R) =$ 1.6--2.2.  
An initial estimate
of the modal colour, its amplitude and the width of the distribution,
$\sigma_{PEAK}$, is made and fed in as the input to the gaussian
fitting routine.  The routine outputs new estimates of these values 
which are iteratively returned to the fitting routine until convergence 
is achieved.  The resultant values are checked by visually inspecting 
the fit.  The gaussian fits generate the colour of the CMR at the 
fiducial magnitude and a measure
of the width of the CMR, together with associated errors, in each radial 
interval.  The fits are presented in Table~\ref{tab:centiles} and illustrated
in Figure~\ref{fig:combhists}.  
The peak of the CMR in the combined cluster evolves bluewards with 
radius at a rate of $d(B-R) / d r_p = -0.022 \pm 0.004$ (where $r_p$ is 
the projected radius from the cluster core), equivalent to $\Delta (B-R) 
\sim -0.11\pm 0.05$ from the central radial bin out to 6\,Mpc.  Beyond 
6 Mpc the constraints on the colour gradient trend of the CMR become less 
reliable and so we limit our discussion to the region within 6 Mpc.

\subsection{The colour--local density relation}

%
%
\begin{figure*} 
\begin{centering}
\psfig{file=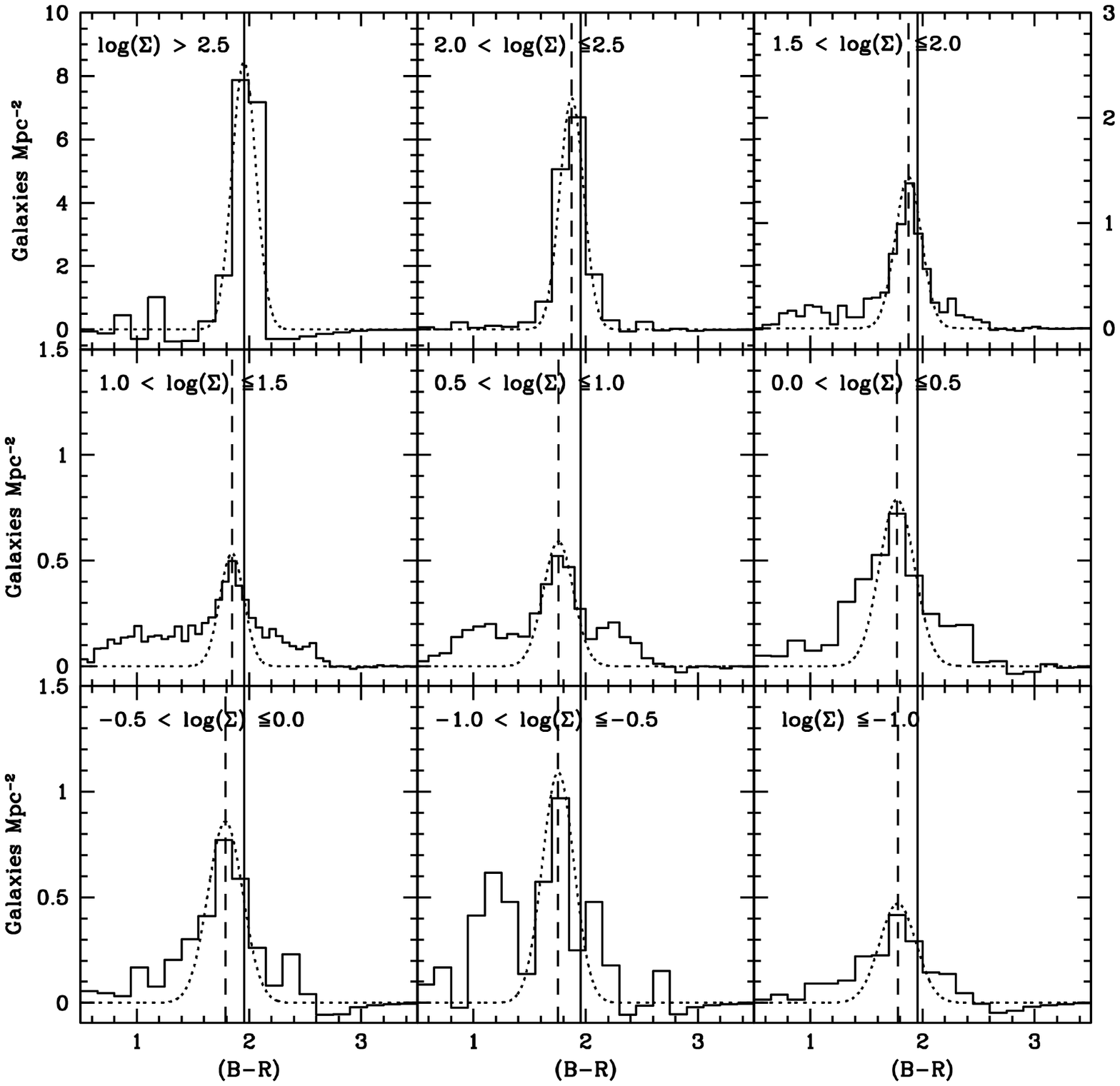,width=17cm} 
\caption{\small{Colour histograms of the composite cluster colour
magnitude relations in logarithmic density intervals after 
correcting for the slope in the CMR and field
contamination.  The best-fit gaussians
are shown for the CMR peaks in each density bin.  The solid vertical
line shows the colour of the peak of the CMR in the highest density bin.
The dashed vertical line is the peak in the particular density interval.
The CMR peak evolves steadily blueward at a rate of  $d(B-R) / d {\rm 
log}_{10}(\Sigma) = -0.076 \pm 0.009$ across three orders of magnitude 
in local galaxy density ($\Sigma$). }}
\label{fig:comb_rho}
\end{centering}
\end{figure*}

As the clusters in our survey exhibit a wide range in their morphologies
(see Figure~\ref{fig:maps}), we seek a more general method of combining
the galaxy samples from the different clusters, rather than simple
radial-averaging, to investigate the environmental differences within
the composite sample.  Local galaxy density provides a comparable
measure of environment across the sample and so we employ this instead.

We calculate the local projected density for each galaxy in our clusters
and use it to assign the galaxy to its relevant environmental bin.
The local galaxy density, $\Sigma$, is estimated by finding the surface
area on the sky occupied by that galaxy and its ten closest neighbours
down to the fiducial magnitude limit of $M_V=-20$.  Due to field galaxy
contamination, this local density will be over-estimated.  We correct
for this by subtracting off a constant density estimated from the median
local density of our combined field sample.  We divide our sample up
into eight logarithmically-spaced bins covering three orders of magnitude
in local galaxy density (Table~\ref{tab:centiles}).

For each local density bin a colour histogram is created.  As before,
the clusters are transformed to a median redshift of $z=0.12$ and the
slope of the CMR is corrected for.  Since these colour histograms are
not corrected for field contamination, it is necessary to subtract off
an appropriately area-scaled colour distribution for our field sample
(see the lower-right panel in Figure~\ref{fig:combhists}).
Finding the area from which a restricted range of densities are drawn 
is non-trivial; a region of high density will have a much smaller field
correction than a region of low density.  
We generate an adaptive map of the field region by binning up 
the positions of the field galaxies and taking the median density
of each bin.  The map is thresholded using density cuts equal to
the eight logarithmically-spaced bins in local galaxy density to 
generate the area occupied by galaxies of that local galaxy density.
This measure of the area is used to scale the colour distribution of
the field sample and subtract it from each local density bin's colour 
histogram.  

The results of this analysis are presented in Figure~\ref{fig:comb_rho}.
Our combined cluster evolves bluewards with decreasing local density
at a rate of $d(B-R) / d {\rm log}_{10}(\Sigma) = -0.076 \pm 0.009$, 
equivalent to $\Delta (B-R) \sim -0.20\pm 0.06$ from the highest local density
regime covered by our sample ($\log_{10}(\Sigma) > 2.5$) to three orders
of magnitude lower.  The results from this analysis provide a 
comparable estimate of the shift in the CMR colour to that identified
in \S4.1 (for the typical cluster in our sample a local galaxy density
of $\log_{10}(\Sigma) \sim 0.5$ corresponds to a radius of $\sim 2$\,Mpc).

\subsection{Width of the CMR}

\begin{figure*} 
\centerline{
\psfig{file=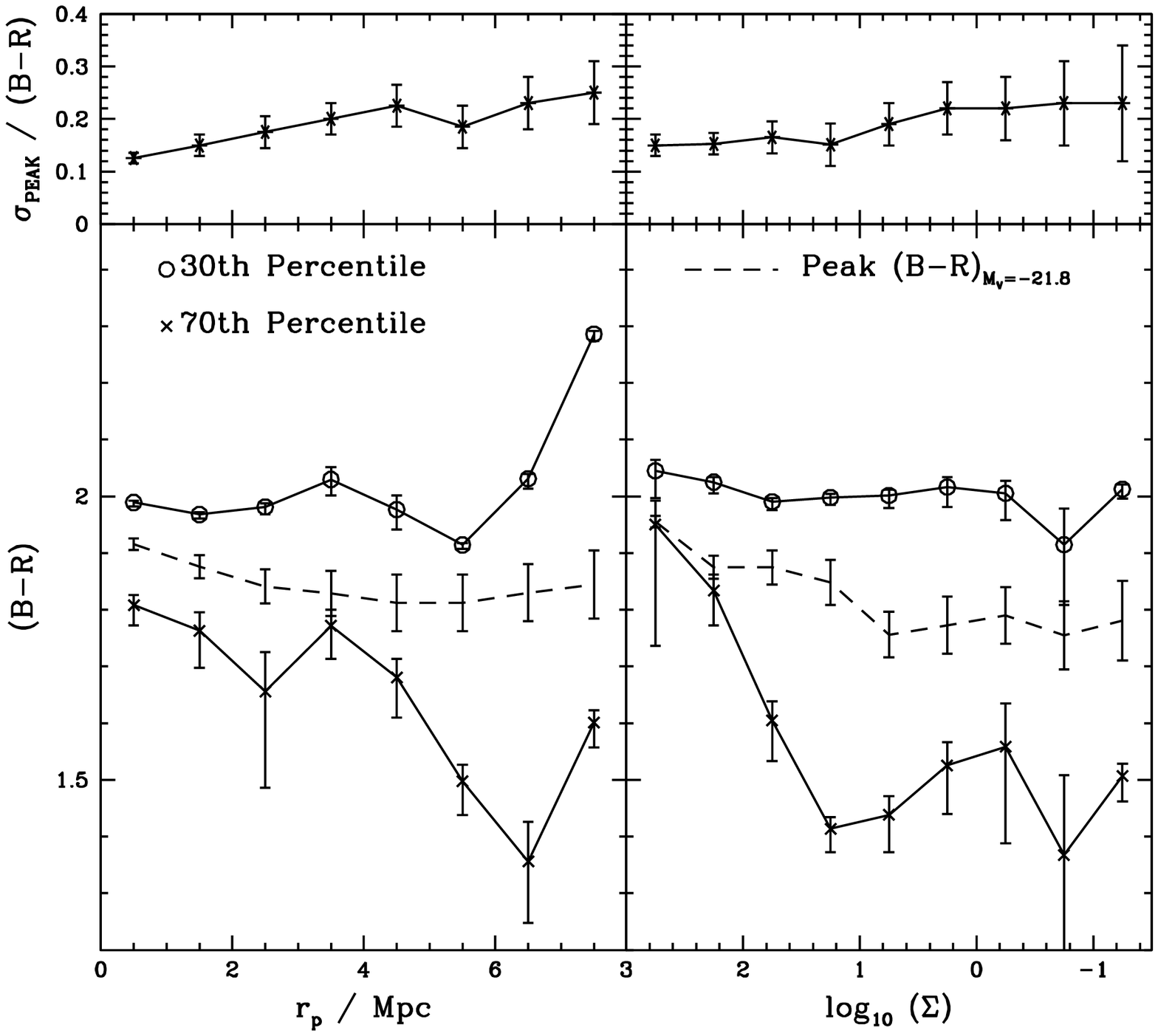,width=15cm,height=14cm} 
}
\caption{\small{The 30$^{\rm th}$ and 70$^{\rm th}$ percentiles of the
colour histograms presented in Figures 5 \& 6, measured relative to the
red end of the distribution.  These are plotted in the two lower 
panels as a function of radius and local galaxy density (solid lines) 
together with the peak CMR colour (dashed line)
taken from Figure~\ref{fig:combhists}.  The 30$^{\rm th}$ percentile of
the distributions is near constant over at least the inner 6\,Mpc of
our combined cluster whilst the peak shifts to the blue by $-0.11\pm
0.05$ mag.  
We suggest that the sharp redward trend seen at large
radii may result from the presence of background clusters and groups
which dominate the red wing at the largest radii.
The upper panel displays the width of the gaussian,
$\sigma_{\rm Peak}$, fit to the colour histograms against radius.  The
width steadily increases by $\sim 0.15$ mag across the radial range.
Similar behaviour is seen in the distributions when sorted in terms
of local galaxy density.}}
\label{fig:centiles}
\end{figure*}

~From the histograms presented in Figures~\ref{fig:combhists} and
\ref{fig:comb_rho}, we now examine how the width of the CMR varies with
environment.  The gaussians fitted to these data in the figures are
used to obtain $\sigma_{\rm Peak}$ which we use as an estimate of the
CMR's width. These values are plotted in the upper panels of
Figure~\ref{fig:centiles}, with tabulated values presented in
Table~\ref{tab:centiles}.  We find that the CMR peak does appear to
broaden with increasing radius and decreasing local galaxy density by
$\Delta \sigma_{\rm Peak} (B-R)\sim 0.15$ across the ranges studied, 
although there is considerable uncertainty in the individual measurements.

%
%
\begin{table}
\begin{center}
\caption{\small{
Peak colour and full width of the CMR's variation with radius and local
galaxy density as estimated from the fitting gaussians to the colour
distributions (Figures 5 \& 6).
}}
\begin{tabular}{ccc}   
\hline
Sample	& Peak $(B-R)_{M_V=-21.8}$ & $\sigma_{\rm Peak}$ \\ \hline \hline
\multispan2{Radius~(Mpc) \hfil }\\	
\noalign{\smallskip}
0--1		& $1.92 \pm 0.01$ & $0.13 \pm 0.01$ \\
1--2		& $1.88 \pm 0.02$ & $0.15 \pm 0.02$ \\
2--3		& $1.84 \pm 0.03$ & $0.18 \pm 0.03$ \\
3--4 		& $1.83 \pm 0.04$ & $0.20 \pm 0.03$ \\
4--5		& $1.81 \pm 0.05$ & $0.23 \pm 0.04$ \\
5--6		& $1.81 \pm 0.05$ & $0.19 \pm 0.04$ \\
6--7	 	& $1.83 \pm 0.05$ & $0.23 \pm 0.05$ \\
7--8 		& $1.84 \pm 0.06$ & $0.25 \pm 0.06$ \\
\noalign{\medskip}
\multispan2{Log~Density~(Mpc$^{-2}$) \hfil }\\	
\noalign{\smallskip}
$>2.5$ 		& $1.96 \pm 0.01$ & $0.15 \pm 0.02$ \\
2.0--2.5   	& $1.88 \pm 0.02$ & $0.15 \pm 0.02$ \\
1.5--2.0 	& $1.87 \pm 0.03$ & $0.17 \pm 0.03$\\
1.0--1.5 	& $1.85 \pm 0.04$ & $0.15 \pm 0.04$ \\
0.5--1.0 	& $1.76 \pm 0.04$ & $0.19 \pm 0.04$ \\
0.0--0.5 	& $1.77 \pm 0.05$ & $0.22 \pm 0.05$ \\
$-0.5$--0.0 	& $1.79 \pm 0.05$ & $0.22 \pm 0.06$ \\
$-1.0$-- $-0.5$ & $1.76 \pm 0.06$ & $0.23 \pm 0.08$ \\
$<-1.0$ 	& $1.78 \pm 0.07$ & $0.23 \pm 0.11$ \\
\hline
\end{tabular}
  \label{tab:centiles}
\end{center}
\end{table}

To further investigate how the colour distribution of galaxies in the CMR
varies with environment we also calculate the colours of the 30$^{\rm th}$
and 70$^{\rm th}$ percentiles from the red end of the colour
distributions (Figures~\ref{fig:combhists} \& \ref{fig:comb_rho}).
The 30$^{\rm th}$ percentile represents the reddest members of the
CMR peak, the typical colour of this population appears to remain
nearly constant in colour (Figure~\ref{fig:centiles}) out to $\sim
6$\,Mpc (i.e.\ the limit of the visibility of the CMR in
Figure~\ref{fig:combhists}). 
Beyond 6 Mpc, there is a reddening of this colour which we 
believe is due to CMRs of other higher redshift clusters.
A similarly selected sample exhibits the same
near constant trend in colour with local galaxy density.  In contrast,
the trend of the 70$^{\rm th}$ percentile, representing the bluer
members of the CMR, shows a strong shift to the blue with both radius
and local galaxy density.

Thus whilst the colour of the reddest members of the CMR appears to be
relatively constant in different environments, the width of the CMR
peak broadens at larger radii and lower local galaxy density,
reflecting an increasing tail of blue galaxies on the outskirts of the
clusters (see Figures~\ref{fig:combhists}, \ref{fig:comb_rho} \& 
\ref{fig:centiles}).

%
%

\subsection{Blue cluster galaxies}

We now compare the relative proportions of the red and blue galaxies
in the clusters.  To do this we quantify the fraction of non-CMR galaxies
by deriving the Butcher-Oemler
blue fraction (Butcher \& Oemler, 1984), $f_B$.  The blue fraction
is calculated for all of our clusters using the method described in
Butcher \& Oemler's (1984) landmark study.  Briefly, to calculate $f_B$
we use only those galaxies whose magnitude is brighter than $M_V = -20$
and lie within a radius of $R_{30}$: the radius of a circle containing 30
percent of the cluster's projected galaxy distribution.  After background
subtraction, the net number of galaxies in these populations are a measure
of cluster richness, denoted $N_{30}$.  The blue fraction is the fraction
of this population whose rest frame colour is $\Delta(B-V)=-0.2$ bluer
than the fitted CMR.  The concentration of the cluster is defined as
$\log_{10} ( R_{60}/R_{20} )$\footnote{The radius $R_{\alpha}$ is the
radius of a circle containing $\alpha$ per cent of the cluster's 
projected galaxy distribution.} and is a measure of the cluster's central
concentration (Butcher \& Oemler 1978, 1984).

We present in Table \ref{tab:bf} the median value of $f_B$ for the
clusters and the associated values of $R_{30}$, $N_{30}$ and the
concentration.  Also noted are the morphologies of the clusters from
Figure~\ref{fig:maps}.

%
%
\begin{table}
\begin{center}
\caption{\small{Median values for the Butcher-Oemler blue fraction, $f_B$,
taken from 100 realizations of the statistical background subtraction.
The errors on $f_B$ are one standard deviation from the median.
Values of $R_{30}$, $N_{30}$ and the concentration, Conc, (Butcher 
\& Oemler 1979) derived from the calculation of $f_B$ are also tabulated.  
The cluster's two-dimensional distribution of CMR members is also 
qualitatively described: Regular, R; Irregular, I.  
}}
\begin{tabular}{lccccc}   
\hline
Cluster & $f_B$ & $R_{30}$ (Mpc) & $N_{30}$ & Conc  & Morph.\ \ \\ 
\hline \hline
A\,22	 & 0.02$\pm$0.02 &  1.06 & 44 & 0.29 & R \\
A\,550	 & 0.04$\pm$0.01 &  1.35 & 96 & 0.32 & R  \\
A\,1079	 & 0.00$\pm$0.05 &  1.32 & 20 & 0.24 & I \\
A\,1084	 & 0.05$\pm$0.03 &  1.78 & 37 & 0.27 & I \\
A\,1285	 & 0.03$\pm$0.01 &  1.40 & 92 & 0.30 & R \\
A\,1437	 & 0.19$\pm$0.02 &  1.11 & 47 & 0.33 & R \\
A\,1650	 & 0.00$\pm$0.04 &  0.52 & 27 & 0.31 & R \\
A\,1651	 & 0.00$\pm$0.02 &  1.19 & 50 & 0.29 & R \\
A\,1664	 & 0.07$\pm$0.03 &  1.20 & 29 & 0.28 & I \\
A\,2055	 & 0.06$\pm$0.03 &  1.26 & 36 & 0.29 & R \\
A\,3888	 & 0.21$\pm$0.02 &  1.01 & 63 & 0.36 & R \\
\hline
\end{tabular}
  \label{tab:bf}
\end{center}
\end{table}

We stress that these values of $f_B$ are likely to be slightly low
due to the biases in the statistical background correction
technique employed.  See Appendix~A for discussion on this point.
There may also be biases arising from the field sample we use.  
Since they are taken at the very edge of the cluster they may be 
fractionally more dense than the `average' field thus
further underestimating $f_B$.  We consider this effect to be small and the
statistical background correction to be the dominant source of
any bias.

There is considerable scatter in the value of $f_B$ in the 
clusters in our sample.
In the majority of our clusters the blue fraction is negligible
and, combined with the small range in X-ray luminosity and redshift
covered by our sample, it is difficult to distinguish any
trends in these parameters.  There are two clusters, 
however, which do have significant blue populations:  A\,1437 and
A\,3888.  Both of these clusters show evidence for recent merger events
(Edge, priv.\ comm.) and appear highly concentrated in X-ray.
The large scatter in $f_B$ for the clusters in our homogeneous sample
is surprising and suggests that this parameter is sensitive to short
term events which affect the galaxy mix within the clusters.

%
%

\section{Discussion}

We have examined variation of the CMR with environment; both with
radius and with local galaxy density.  We find that the CMR peak
evolves bluewards and broadens with both radius and local galaxy density
(Figures~\ref{fig:combhists},~\ref{fig:comb_rho} \&~\ref{fig:centiles}).
Yet the colour of the reddest members of the CMR remains constant
(Figure~\ref{fig:centiles}).  
We emphasize at the outset of our discussion that the effects of 
environment upon 
galaxy evolution (e.g. star formation suppression, triggering and 
the morphological evolution of galaxies) are still polemical issues 
in modern astronomy (Dressler et al.\ 1997; Lubin et al.\ 1998; Andreon 1998;
van Dokkum et al.\ 1998).   
In this section, we examine possible causes
for these effects with particular reference to the S0 evolution
proposed by the MORPHS team (e.g. Dressler et al.\ 1997; Poggianti et al.\ 1999).

\subsection{The effects of morphology on the CMR}

We now estimate the trends which should exist in our sample due to the
existance of a morphology-radius relation (T--R, Whitmore \& Gilmore
1991; Whitmore et al.\ 1993) or a morphology-density 
relation (T--$\Sigma$, Dressler 1980).
We adopt a null hypothesis that the colour gradient we observe in the
CMR with environment is due to an increasing fraction of early-type spiral
galaxies in the cluster sample, while the colours of the spheroidal
galaxies (ellipticals and S0s) are unaffected by their environment.
This hypothesis is suggested by the fact that we see a broadening of
the colour-magnitude relation with radius, but a fixed red envelope
(Figure~\ref{fig:centiles}).

Although we really do not have detailed morphological classifications
for our sample,
we can use the distribution of galaxies in terms of concentration
index (CI) and peak surface brightness ($\mu_{\rm MAX}$) to look for
changes in the morphological mix in different environments within the
clusters.  On the CI--$\mu_{\rm MAX}$ plane, early-type galaxies populate the
high-concentration and high surface brightness region, with later-types
typically having lower concentrations and fainter peak surface
brightnesses (Abraham et al.\ 1994; Pimbblet et al.\ 2001).  Comparing
the distribution in CI--$\mu_{\rm MAX}$ of galaxies lying on the CMR in
the cluster core ($r_p < 1$\,Mpc) and outskirts ($r_p>4$--6\,Mpc) we
also find evidence for a shift in the typical morphology  -- with the
outer regions having a larger proportion of lower-concentration,
lower-surface brightness galaxies than the core.  To quantify this
difference, we use a two dimensional Kolmogorov-Smirnov (K-S) test
(Fasano \& Franceschini 1987) to show that the CI--$\mu_{\rm MAX}$
distribution for the  $r_p < 1$\,Mpc and $r_p>4$--6\,Mpc samples are
unlikely to be drawn from the same parent population at 97.5 per cent
confidence.  Unfortunately, it is non-trivial to transform the
observed shift in the CI--$\mu_{\rm MAX}$ distribution between these
environments into expected colour differences.

There is another approach we can use to test our null hypothesis: 
we can evaluate the typical $(B-R)$ colour of the
galaxy distribution in the clusters based on  Whitmore \& Gilmore's
(1991) radial morphological mix and Dressler's (1980) local-density
morphological mix.  We adopt the morphological mix in local rich
clusters on the assumption that it is unlikely to 
change dramatically from $z=0$ to
$z=0.12$ (Fasano et al.\ 2001; Dressler et al.\ 1997).  
Our limiting magnitude of $M_V=-20$ is also broadly comparable with
Dressler's (1980) limit of $M_V \sim -20.4$.
We use the
$(B-V)$ colours of the different morphological types from the Third
Reference Catalogue  (RC3, de Vaucouleurs et al.\ 1991) as illustrated
in Roberts \& Haynes (1994) and assume a no-evolution K-correction
based on the SEDs in King \& Ellis (1985).  We split the sample into
broad morphological types corresponding to those employed by Dressler
(1980) and Whitmore \& Gilmore (1991): E, S0 and S+Irr.  The $(B-V)$
colours of E and S0 galaxies from the RC3 are $(B-V)=0.90$ and 0.89
respectively.   These correspond to observed $(B-R)$ colours at
$z=0.12$ of $(B-R)=1.94$ for E's and $(B-R)=1.91$ for S0s.  The third
morphological class is ``S+Irr'':  comprising spiral and irregular
galaxies.  The use of this class in our analysis, however, is
problematic due to its broadness.  The CMR will not contain many of the
later-type spirals (e.g.\ Sc and Sd) as their mean colours  lie
significantly bluewards of the CMR defined by elliptical and S0
galaxies.

To examine which morphological types from the ``S+Irr'' class are
likely to inhabit the CMR we estimate how blue a galaxy has to be
before it falls outside of the CMR peak and hence does not affect our
gaussian fitting to the colour distributions
(Figures~\ref{fig:combhists} \&~\ref{fig:comb_rho}).  We estimate
that galaxies as blue as $(B-R)\sim 1.5$--1.6 will affect the fitting
of the CMR in the outskirts of the clusters; this corresponds to
$(B-V)\sim 0.7$ at $z=0$ or the colour of an Sab galaxy from RC3.
At $z=0.12$, an Sa galaxy will have a mean colour of $(B-R)= 
1.55$ while Sab's have $(B-R)= 1.49$.
Hence, we expect that it is only the Sa/Sab class from the spiral
population which is likely to influence the colour of the CMR based on
our fitting procedure.  We therefore use Sa/Sab as our third
morphological class, instead of the broader ``S+Irr'' class.  
Before we calculate the expected colour of the CMR peak we re-normalize
the T--R and T--$\Sigma$ morphological mixtures to account for using
only Sa/Sab galaxies.  This again is done using the RC3, from which we
estimate the fraction of Sa/Sab galaxies within the ``S+Irr'' class and
then re-scale the relative proportions of the three morphological
classes (Table~\ref{tab:radcols}).

%
%
\begin{table}
\begin{center}
\caption{\small{Relative changes in the colours of the cluster
population at $z=0.12$, compared to the galaxy population in the
cluster core, assuming the morphological mixes from Whitmore \& Gilmore
(1991) and Dressler (1980).  A blueward shift   of $-$0.07 mags in
$(B-R)$ is predicted from the centre out to 4\,Mpc and $-$0.08 mags
over a two orders of magnitude change in local density from the core.
}}
\begin{tabular}{ccccc}   
\hline
Sample  & E & S0 & Sa/Sab & $\Delta(B-R)_{M_V=-21.8}$ \\ \hline \hline
\multispan4{Radius~(Mpc) \hfil } \\
\noalign{\smallskip}
0--1 	& 0.27 & 0.67 & 0.06 & ~~0.00 \\
1--2 	& 0.25 & 0.60 & 0.15 & $-$0.04 \\
2--3 	& 0.23 & 0.59 & 0.18 & $-$0.05 \\
3--4 	& 0.21 & 0.56 & 0.23 & $-$0.07 \\
\noalign{\medskip}
\multispan4{Log~Density~(Mpc$^{-2}$) \hfil }\\	
\noalign{\smallskip}
$>$2.5	 & 0.45 & 0.55 & 0.00 &  ~~0.00 \\
2.0--2.5 & 0.43 & 0.53 & 0.04 & $-$0.02 \\
1.5--2.0 & 0.36 & 0.55 & 0.09 & $-$0.04 \\
1.0--1.5 & 0.29 & 0.56 & 0.15 & $-$0.07 \\
0.5--1.0 & 0.21 & 0.59 & 0.20 & $-$0.09 \\
0.0--0.5 & 0.20 & 0.57 & 0.23 & $-$0.10 \\
\hline
\end{tabular}
  \label{tab:radcols}
\end{center}
\end{table}

The gradient in mean $(B-R)$ colour predicted from the variation in
morphological mix seen in the T--R relation (Table~\ref{tab:radcols})
is $d(B-R) / d r_p = -0.017$.  This is equivalent to a change of 
$\Delta(B-R)=-0.07$ out to 4\,Mpc which is roughly 75 percent of
that observed in the combined LARCS clusters ($d(B-R) / d r_p = 
-0.022 \pm 0.004$).
The gradient in the CMR colour resulting from the variation
in the morphological mix seen in the T--$\Sigma$ relation is
$d(B-R) / d {\rm log}_{10}(\Sigma) = 0.033$.
This is equivalent to a change from
$\log_{10}(\Sigma)>2.5$ to $\log_{10}(\Sigma)=0.0$ of $\Delta(B-R)=-0.10$,
which is around 50 percent of that seen over this density range in
the composite of the LARCS clusters ($d(B-R) / d {\rm log}_{10}(\Sigma) 
= -0.076 \pm 0.009$).
We find similar predicted trends at 99 per cent confidence
if we restrict ourselves to only the high $L_X$
clusters from Dressler (1980). 

\subsection{Interpretation}

Based on the results from the previous section we conclude that the
morphology-radius and morphology-density relations seen in local
clusters can account for only around half of the apparent change in
galaxy colour with radius and local galaxy density seen in our sample.

We estimate therefore that there is an {\it intrinsic} gradient corrected for 
morphological differences in the colours of early-type galaxies 
at a fixed luminosity of $d(B-R) / d {\rm log}_{10}(\Sigma) = -0.011$
from the cores of typical rich clusters at $z\sim 0.12$ 
out to environments with galaxy densities of $\log_{10}(\Sigma)\sim 0$
(equivalent to a radius of 2\,Mpc).  The intrinsic variation with 
clustocentric radius gives a somewhat weaker trend (Table~\ref{tab:radcols}):
$d(B-R) / d r_p = -0.005$.

Radial gradients in the colours of cluster galaxies have been
previously  reported by Abraham et al.\ (1996) and Terlevich et
al.\ (2001) for two rich clusters straddling the redshift range studied
here.  These two measurements are not exactly comparable as the
Terlevich et al.\ analysis focuses on the morphologically classified E
and S0 galaxies, whereas Abraham et al.\ investigate a colour-selected
sample around the CMR, similar to the approach used here.
Nevertheless, as Terlevich et al.\ stress, the core of Coma is
completely dominated by E and S0 galaxies and so their results would
not change significantly if they considered the whole population.

Abraham et al.\ (1996) find radial gradient in $(g-r)$ colour of
$d(g-r)/d\log_{10}(r_p)=-0.079$ out to 5\,Mpc in A\,2390 ($z=0.23$),
this is not corrected for any variations due to a $T-R$ relation.  
We estimate there is
$\sim 20$ percent uncertainty in this gradient.  Terlevich et
al.\ (2001) report $d(g-r)/d\log_{10}(r_p)=-0.024\pm 0.005$ for the Coma
cluster, transformed to the same restframe wavebands as the
observations A\,2390.  To compare our result with these we similarly
transform the $(B-R)$ colours of galaxies in our composite $z=0.12$
cluster to the wavelengths equivalent to observing a cluster in $(g-r)$
at $z=0.23$.  The result of this transformation is a radial gradient of
$d(g-r)/d\log_{10}(r_p)=-0.061\pm 0.011$ (uncorrected for the 
$T-R$ relation).  Hence, the radial gradient of
the colour of the CMR at $z=0.12$ (a look-back time of 2\,Gyrs) is
intermediate between those found by Abraham et al.\ (1996) for A\,2390
and Terlevich et al.\ (2001) for the Coma cluster at look-back times of
3.5\,Gyr and 0.5\,Gyrs respectively.
Correcting the higher redshift measurements for the effects of the 
morphology-radius relation still results in steeper gradients
at higher redshifts.

Both Abraham et al.\ (1996) and Terlevich et al.\ (2001) attribute the
colour gradients they observe in the clusters to age trends (Terlevich
et al.\ 2001 use detailed arguments to show that the variation in
colour cannot be due to dust spread through the core of the cluster).
Using the models for a single burst stellar population
at an age of 7\,Gyr (equivalent to the time elapsed between a formation
epoch at $z\geq 2$ and $z=0.12$) presented in Bower et al.\ (1998), we
find $d(B-R) / dt \sim 0.03$ mag\,Gyr$^{-1}$ (Kodama \& Arimoto, 1997).
This implies that the galaxies on the CMR in the outer regions of our
clusters have luminosity-weighted stellar populations which are
approximately 3\,Gyrs younger than those in the cores ($r_p < 1$\,Mpc).
This age difference is comparable to that found by Terlevich et
al.\ (2001) who predict that early-type galaxies in the outskirts of
Coma have stellar populations which are $\sim 2$\,Gyr younger than
those in galaxies in the central regions.
 
We also propose that the colour gradients observed in our sample
represent variations in the ages of the stellar populations in the
cluster galaxies arising from diffences in their star formation
histories (Abraham et al.\ 1996; Smail et al.\ 1998, 2001; Kodama \&
Bower 2000).  The bulk of galaxies on the CMR are expected to be S0's
(Table~\ref{tab:radcols}; Dressler 1980; Fasano et al.\ 2001) and thus
changes in the colours of this morphological class must play a major
role in causing the observed colour gradient (Abraham et al.\ 1996).
However, we also see evidence for a proportion of old, evolved galaxies
whose colours are constant with environment and define the red wing of
the CMR (Figure~\ref{fig:centiles}).  We suggest that the majority of
these galaxies are probably ellipticals, which show modest
environmental variation in their properties (Treu et al.\ 1999).  
Thus a mix of ellipticals with constant colours and a dominant population 
of S0 galaxies, whose colours vary with environment, can reconcile the 
main results from our survey.

Why should galaxies which have S0 morphologies, or at least the 
colours of S0's at the present-day, vary with environment?  
Although still controversial,
there is increasing evidence that the S0 population in 
rich clusters has come
into being relatively recently. The proportion of S0 galaxies in rich
clusters appears to increase rapidly to the present-day (Dressler et al.\ 1997;
Fasano et al.\ 2001), possibly as the result of the transformation of
star-forming disk galaxies which are accreted by the clusters from the
surrounding field (Poggianti et al.\ 1999;  Kodama \& Smail 2001).  The
fact that we can see differences in the properties of cluster galaxies
out at least as far as 6\,Mpc from the cluster core suggests that
the process which may suppress the star formation of infalling
galaxies can operate in low density environments (Fasano et al.\ 2000;
Bekki et al.\ 2001; Carlberg et al.\ 2001; Couch et al.\ 2001; Kodama 
et al.\ 2001).  One 
mechanism for causing this transformation is the suppression of star
formation due to the removal of the gas reservoirs as galaxies enter
the cluster (Poggianti et al.\ 1999; Balogh, Navarro \& Morris 2000).
This mechanism may lead directly to the formation of an S0, or other
processes may be involved (Balogh et al.\ 2001).

If this scenario is correct then the accretion and transformation of
spiral galaxies from the surrounding field will lead to a gradual
build-up of the S0 population within the cluster.  Some morphological
studies of the early-type galaxies in cluster cores at $z\sim 0.5$
imply that the majority of the S0 population in these regions today are
transformed in the 6\,Gyrs between $z=0.5$ and the present-day
(e.g. Poggianti et al.\ 1999).  On the
outskirts of the clusters we expect a higher proportion of recently
arrived cluster members, reflecting the accretion history of the
cluster (Abraham et al.\ 1996).  Thus the S0 galaxies in these regions
will show the strongest signatures of their past star formation
activity, exhibiting the youngest luminosity-weighted ages for their
stellar populations and hence the bluest colours.  These two effects
will combine to produce a radial colour gradient in the same sense as
that seen in our data.  After accounting for the varying morphological
mix with environment, it appears that galaxies on the outskirts of
these clusters have luminosity-weighted ages which are $\sim 3$\,Gyrs
younger than those in the core.  It would be interesting to
quantitatively compare this observation with theoretical predictions to
test models for the infall history and transformation of galaxies in
high density environments (Diaferio et al.\ 2001).

\subsection{Caveats}

Our analysis and interpretation
is based upon a number of assumptions that require consideration
and should be improved upon in future work.
Perhaps the largest source of uncertainty is
the background subtraction method employed.   It is apparent
from Figure~\ref{fig:inners} that there are a number of galaxies, which 
are rejected by the background subtraction algorithm, but which lie 
within the CMR (e.g. Abell~1650).  Conversely, there
must also be galaxies which are defined as cluster
members, but which are in all likelihood interlopers.
We re-emphasize that the background correction is 
purely a statistical method.
There are two clear ways in which this method could be improved.
Firstly, we could use a larger sample of 
dedicated field observations to improve the statistics.  
Secondly, we could use spectroscopic observations of the
clusters to provide a more robust method to determine cluster 
membership.

During our analysis we combine our clusters into a single
composite cluster to examine the radial dependance of the
CMR.
We note that the richer clusters (i.e. Abell~550,~1285 and~3888)  
have a greater weight in the composite cluster than 
the poorer ones (i.e. Abell~1079,~1084 and~1664).
Although cluster-to-cluster variations are present within our
sample, we emphasize that LARCS represents a 
homogeneously X-ray selected sample of clusters and therefore
such cluster-to-cluster variations should be minimized.
Using more clusters to create the combined cluster would
be beneficial.

In our discussion we compare our data to that of Dressler (1980)
to remove the $T-\Sigma$ relation from our clusters.
We note that the LARCS dataset does not contain sufficiently
high resolution imaging required morphologically classify galaxies.
Hence we are unable to derive a
$T-\Sigma$ relation for our clusters and we rely upon the morphology-density 
analysis performed by Dressler (1980) for other clusters.
In using these data, we note that the Dressler (1980) sample 
is constructed with an inhomogeneous local ($z<0.06$) sample of 
clusters and may, therefore, not be directly comparable to our 
study (although limiting Dressler's sample only to high $L_X$ 
clusters gives the same results).  
To remedy this situation would require sub-arcsecond panoramic imaging
of our clusters to provide the necessary morphological information.

%
%
\section{Conclusions}

We have analysed eleven clusters with precise photometric observations
from the LARCS survey to trace the variation in the colours of evolved
galaxies into the environmental transition region between clusters
and the field.

\smallskip

\begin{itemize}

\item All of the clusters show colour-magnitude relations; many of
these are strong and can be traced well beyond the central 1\,Mpc,
out to unprecedented clustocentric radii coverage of $\sim 8$ Mpc.

\item We find that the characteristic colour and slope of the CMR of
galaxies in the cluster cores change with cluster redshift across the
range $z=0.07$--0.16 covered by our sample.  The variations in the CMRs
are compatible with those expected from galaxy populations which
formed the bulk of their stars at high redshifts, $z\gg 2$.

\item The majority of clusters in the survey exhibit a variation with
radius in the colour of the CMR at a fixed luminosity.  When combined
into a single composite cluster, we observe a radial colour gradient of $d(B-R) 
/ d r_p = -0.022 \pm 0.004$ from the cluster centre to $\sim 6$\,Mpc.

\item The composite cluster formed from our sample also shows a
colour-local-density relation for the galaxies on the CMR.  We identify
a colour gradient in the modal colours of galaxies on the CMR amounting to
$d(B-R) / d {\rm log}_{10}(\Sigma) = -0.076 \pm 0.009$  across three 
orders of magnitude in local galaxy density from the high-density core
of the cluster to the outskirts.

\item We calculate the expected change in the colour of the CMR with
environment due to the changing morphological mix.  We estimate that
approximately 50 to 70 per cent of the radial and density colour 
gradient is due
to an increase in contamination by quiescent, early-type spiral 
galaxies falling close to the CMR.  We suggest that the CMR 
consists of a passively evolving population (elliptical and evolved 
S0 galaxies) plus recent additions (morphologically young S0 and 
recently star-forming spiral galaxies).
  
\item Our analysis of eleven clusters supports the previous claims of
radial colour variations in individual rich clusters at $z=0$ and
$z=0.23$.  We have also shown that similar trends are visible when
using local galaxy density as a more general measure of galaxy
environment.  Taken together with previous studies, we suggest that
these environmental trends in the colours of galaxies at a fixed
luminosity most likely reflect differences in the luminosity-weighted
ages of the galaxies in different environments.  If interpreted purely
as a difference in ages and accounting for differences in the 
morphological mix, then the gradient observed in our sample
suggests that the luminosity-weighted ages of the dominant galaxy
population within the CMR at 6\,Mpc from
the cluster core are $\sim$3\,Gyrs younger than those residing in the core.

\item The results of our photometric analysis can provide a readily
testable, quantitative prediction.  If the observed colour gradient
simply reflects age differences in passive stellar populations then we
would expect that the typical  $H\delta_A + H\gamma_A$ line strength of
galaxies lying on the CMR should increase by $\sim 4$\AA\ when moving
from the cluster core to the outskirts of these systems (Terlevich et
al.\ 1999).

\item Further studies are urgently needed to investigate the properties
of galaxies in the lower density environments which connect rich
clusters to the surrounding field, particularly at higher redshifts.  
This region is essential for investigating the impact of environment 
upon the star formation and morphologies of galaxies.

\end{itemize}

\smallskip

This is the second paper in a series based upon the LARCS 
survey.  In the next paper we will examine 2dF spectroscopy for
several of the clusters from the survey (O'Hely et al.\ in prep.).

\subsection*{Acknowledgements}

We thank Richard Bower and Peter S. Craig for 
insightful and stimulating discussions.  
We also thank Michael Balogh, Ray Sharples and Simon Morris for their 
careful reading of this manuscipt and providing many useful suggestions
which have improved this work.
KAP acknowledges support from his PPARC studentship.  IRS
acknowledges support from the Royal Society and a Philip Leverhulme
Prize Fellowship.  TK acknowledges the Japan Society for Promotion of
Science through its Research Fellowships for Young Scientists.  WJC 
and EO'H acknowledge the financial support of the Australian Research 
Council.  ACE acknowledges support from the Royal Society.  AIZ
acknowledges support from NASA grant HF-01087.01-96.  We thank the
Observatories of the Carnegie Institution of Washington for their
generous support of this survey.  This research has made extensive use
of the University of Durham's {\sc STARLINK} computing facilities,
expertly managed by Alan Lotts, and the facilities at Las Campanas
Observatory, Chile.

%
%

\medskip

\section*{Appendix A: Statistical Correction.}

To examine the cluster galaxy population in the absence of
spectroscopy, it is necessary to statistically subtract the field
population from a given cluster plus field population.  Typically, to
obtain the cluster population one subtracts off an area and
density-scaled portion of the defined field population, $F_{\rm
pop}$, from the defined cluster plus field population, $C+F_{\rm
pop}$.  Both the $F_{\rm pop}$ and the $C+F_{\rm pop}$ are then binned
onto a sensibly gridded colour-magnitude diagram.  A straight forward
subtraction is then performed to calculate the probability of a galaxy
in a given colour-magnitude grid position of being a field galaxy.
Thus: 
\begin{equation}
  P(\rm Field)_{\rm col, mag} 
= \frac{N(\rm Field)_{\rm col, mag} \times A}{N(\rm Cluster+Field)_{\rm col, mag}}
\end{equation}
where $A$ is an area- and density-dependant scaling factor.  For each
galaxy in $C+F_{\rm pop}$, we proceed to generate a random number
between 0.0 and 1.0 and compare it to the galaxies' value of $P(\rm
Field)_{\rm col, mag}$ in order to determine if the given galaxy should
be in the resultant cluster population.  Problems in this method occur
when directly subtracting the populations on the color-magnitude plane
produces a negative number of galaxies in the cluster population, $C_{\rm popn}$, for a particular grid position.  A resultant
$P(\rm Field)_{\rm col, mag}$ can thus be greater than 1.0.  Therefore
to utilize a background subtraction method to analyze a cluster
population in the absence of spectroscopy, it is necessary to solve
this ``negative galaxy'' problem.  A typical method 
is to normalize all of the probabilities to the largest value of $P(\rm
Field)_{\rm col, mag}$.  Clearly such a solution has drawbacks as
the resultant $P(\rm Field)_{\rm col, mag}$ distribution no longer
reflects the appropriately scaled input data and has actually been
scaled away from its true value in many locations on the
color-magnitude plane for the sake of one (or a few) anomalously low
counts in $C+F_{\rm pop}$ for a particular grid position.

The method adopted in this study is similar to that used by Smail et
al.\ (1998) and Kodama \& Bower (2001, KB01).  If, when gridding up the
colour-magnitude diagrams, we produce a negative galaxy  in the
computation of $P(\rm Field)_{\rm col, mag}$, we solve it by increasing
the grid size {\it for that particular grid position}.  Thus the grid
position $({\rm col}, {\rm mag})$ is increased to include its nearest
neighbours $({\rm col}:{\rm col}+1, {\rm mag}:{\rm mag}+1)$ and then
$P(\rm Field)_{\rm col, mag}$ is recalculated using this new $2\times2$
grid.  If this is not enough to cure the negative galaxy problem, we
increase the grid size for that particular grid position by one further
step to $({\rm col}-1:{\rm col}+1, {\rm mag}-1:{\rm mag}+1)$ and so on
until the field-corrected galaxy count exceeds zero.

\begin{figure*} 
\begin{centering}
\psfig{file=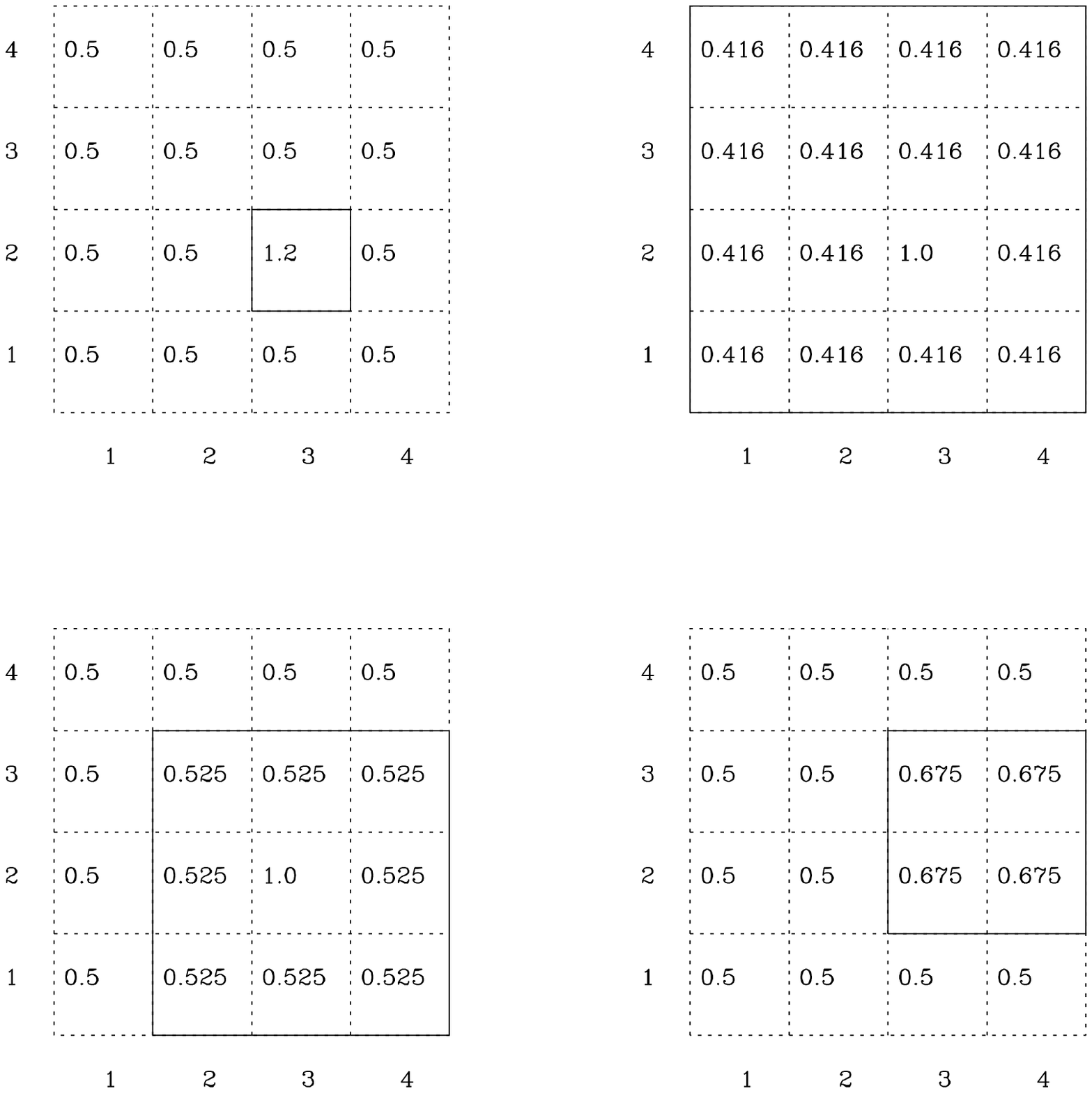,width=6in,height=6in} 
\caption{\small{
A comparison of the statistical correction methods on the
colour-magnitude plane.  The galaxies have been binned into a
colour-magnitude grid and the calculated $P(\rm Field)$ is shown for
each grid position.  [top left]: An example distribution of $P(\rm
Field)$ containing a grid position with $P(\rm Field) > 1.0$ -- the
negative galaxy problem.  [top right]:  Normalizing {\it all} the points
to the anomalous grid position retains the original anomalous point and
artificially decreases the probabilities in the remainder of the grid.
[bottom left]: The approach used by KB01:  the excess 0.2 probability is
distributed equally into neighbouring grid cells.  [bottom right]:
Adaptive-mesh method used in this work -- the lowest possible number of
grid positions are used to recalculate the $P(\rm Field)$ distribution
and no obvious fluctuation remains.
}}
\label{fig:probs}
\end{centering}
\end{figure*}

In Figure~\ref{fig:probs} we compare the resultant $P(\rm Field)_{\rm
col, mag}$ distribution using this method and KB01.  The grid position
$(3,2)$ has a fluctuation in $C+F_{\rm pop}$ so that the number of
galaxies in the corresponding grid position in $F_{\rm pop}$ makes
the resultant $P(\rm Field)_{3,2} > 1.0$.  By normalizing to the
anomalous value, {\it all} the rest of the probabilities are reduced
from their original values and we are still left with a high
probability at the original grid position.  The KB01 method subtracts
the excess probability (in this case $0.2$) and distributes it evenly
between its neighbours, making the original grid position $P(\rm
Field)_{3,2} = 1.0$.  Using our new method, we resample grid positions
$(3:4,2:3)$ to make one new, larger grid position and calculate the new
probability for this.

%
%

It is found that the negative galaxy problem exists in all clusters
examined in this study.  The number of grid-expansions required to cure
the problem are typically around ten $2\times2$ expansions and a small
number of larger order expansions.  The expansions are typically
situated away from the CMR, preferentially where the $C+F_{\rm pop}$
is sparse.

As with all background subtraction methods, this technique has 
advantages and drawbacks.  Whilst it preserves the original probability
distributions better than other methods, it also has a tendency to ``smear''
the resultant distribution.  To illustrate this one can imagine a situation
where one grid position has a $P(\rm Field) > 1$ whilst its neighbours
have very small values of $P(\rm Field)$.  Upon expansion of the grid,
the probabilities of being a field galaxy significantly increase for 
those grid positions that initially had a very low $P(\rm Field)$.  
Therefore we effectively remove galaxies from adjoining grid positions
to make up for the one grid position which had $P(\rm Field) > 1$.
It is because of this that we may expect that statistics on sparsely
populated areas of a colour-magnitude diagram to be adversely affected
(e.g. it is likely that the blue population, and hence $f_B$, 
is underestimated).  Unlike previous methods, this new technique
minimizes the number of grid positions affected in this manner.

\end{document}